\documentclass[aps,prx,10pt,twocolumn,a4paper, nofootinbib, notitlepage, longbibliography]{revtex4-1}

\usepackage{amsmath}
\usepackage{textcomp}
\usepackage{pifont}
\usepackage{upgreek}
\usepackage{units}
\usepackage[pdftex]{graphicx}

\usepackage{txfonts}

\usepackage{hyperref}
\hypersetup{colorlinks=true, citecolor=blue, urlcolor=blue, linkcolor=blue}

\renewcommand{\vec}[1]{\boldsymbol #1}

\usepackage{pifont}

\begin{document}
\title{Interaction-stabilized topological magnon insulator in ferromagnets}

\author{Alexander Mook}
\affiliation{Department of Physics, University of Basel, Klingelbergstrasse 82, CH-4056 Basel, Switzerland}

\author{Kirill Plekhanov}
\affiliation{Department of Physics, University of Basel, Klingelbergstrasse 82, CH-4056 Basel, Switzerland}

\author{Jelena Klinovaja}
\affiliation{Department of Physics, University of Basel, Klingelbergstrasse 82, CH-4056 Basel, Switzerland}

\author{Daniel Loss}
\affiliation{Department of Physics, University of Basel, Klingelbergstrasse 82, CH-4056 Basel, Switzerland}

\begin{abstract}
Condensed matter systems admit topological collective excitations above a trivial ground state, an example being Chern insulators formed by Dirac bosons with a gap at finite energies. However, in contrast to electrons, there is no particle-number conservation law for collective excitations. This gives rise to particle number-nonconserving many-body  interactions whose influence on single-particle topology is an open issue of fundamental interest in the field of topological quantum materials. Taking magnons in ferromagnets as an example, we uncover topological magnon insulators that are stabilized by interactions through opening Chern-insulating gaps in the magnon spectrum. This can be traced back to the fact that the particle-number nonconserving interactions break the effective time-reversal symmetry of the harmonic theory. Hence, magnon-magnon interactions are a source of topology that can introduce chiral edge states, whose chirality depends on the magnetization direction. Importantly, interactions do not necessarily cause detrimental damping but can give rise to topological magnons with exceptionally long lifetimes. We identify two mechanisms of interaction-induced topological phase transitions---one driven by an external field, the other by temperature---and show that they cause unconventional sign reversals of transverse transport signals, in particular of the thermal Hall conductivity. We identify candidate materials where this many-body mechanism is expected to occur, such as the metal-organic kagome-lattice magnet Cu(1,3-benzenedicarboxylate), the van der Waals honeycomb-lattice magnet CrI$_3$, and the multiferroic kamiokite (Fe$_2$Mo$_3$O$_8$). Our results demonstrate that interactions can play an important role in generating nontrivial topology. 
\end{abstract}

\maketitle

% ====================================
%  INTRODUCTION
% ====================================
\section{Introduction}
Over the last decades, the field of solid state physics saw the consolidation of quantum state geometry and topology \cite{Kolodrubetz2017} in Bloch's band theory \cite{Bloch1929}, a prominent result of which is the concept of Chern insulators, exhibiting the quantum anomalous Hall effect due to topologically protected chiral edge states \cite{Klitzing1986, pankratov1987supersymmetric, Haldane1988, Chang2013, Liu2016QAHE}. 
Besides electrons, the collective excitations of crystals also obey Bloch's band theory, e.g., phonons or magnons, the latter being quanta of spin waves in magnetically ordered materials \cite{Bloch1930}. With magnetic properties easily manipulated by magnetic fields, magnonic topology offers a unique external handle to explore the rich and still surprising fundamentals of band theory.
By now, several topological magnon phases have been proposed and material candidates identified; the list includes 
magnon Chern insulators \cite{Meier2003, Katsura2010, Zhang2013, Shindou13, Shindou2013b, Hoogdalem2013, Mook14b, Mook2016spintronics} 
in Cu(1,3-benzenedicarboxylate) \cite{Chisnell2015}, CrI$_3$ \cite{Chen2018}, and YMn$_6$Sn$_6$ \cite{Zhang2020}, 
topological magnon $\mathbb{Z}_2$ insulators \cite{Nakata2017QSHE, Lee2018, Mook2018, Kondo2019, Kondo2019b}, 
Dirac magnons \cite{Fransson2016} in CrBr$_3$ \cite{Yelon1971, Pershoguba2018}, Cu$_3$TeO$_6$ \cite{Li2017,Yao2018, Bao2018}, and CoTiO$_3$ \cite{Yuan2020, Elliot2020arxiv}, 
Weyl magnons \cite{Li2016, Mook2016, Su2016, Su2017, Jian2018, Liu2019MagnonQuantumAnomaly, Zhang2020Weyl} in Lu$_2$V$_2$O$_7$ \cite{Mook2016} and Cu$_2$OSeO$_3$ \cite{Zhang2020Weyl}, 
nodal-line magnons \cite{Mook2016b, owerre2017dirac, Owerre_2019, Liu2020AFMNodal, Elliot2020arxiv}, chiral topological magnon insulators \cite{Li2018ChiralMagnonInsulator}, and second-order topological phases \cite{Li2019HigherTopSolitons, Sil2020, Li2020HigherOrder, Li2020HigherOrderSquare, Hirosawa2020TopMagnQuad, Mook2020ChiralHingeMagnons}. Out of the first-order phases, the magnonic Chern insulating phase is the most fundamental because the others rely on additional symmetries: it takes an effectively fermionic time-reversal symmetry (causing magnonic Kramers partners) for magnon $\mathbb{Z}_2$ insulators, Dirac magnons need the simultaneous presence of an (effective) time-reversal and an inversion symmetry, Weyl and nodal-line magnons are well-defined only for translationally invariant systems, and the chiral symmetry stabilizes chiral insulators. Hence, in principle, arbitrarily small symmetry-breaking perturbations destroy topology. In contrast, the Chern insulating phase is stable against considerable disorder \cite{Ruckriegel2018} (too weak to close the topological band gap), promoting the topologically protected chiral edge magnons to robust, unidirectional information highways, which are free of Joule heating due to the chargelessness of magnons. Hence, the actively studied field of ``topological magnonics'' emerged \cite{Shindou13, Shindou2013b, Mook15a, Mook15b, Molina2016, Wang2018TopMagnonics,Malki2019, Diaz2019FMSkX}.

In contrast to electrons, the absence of a conservation law for the magnon particle number admits  number-nonconserving many-body interactions and spontaneous decays \cite{Zhitomirsky2013}.
Nonetheless, neglecting magnon-magnon interactions is widely assumed in studies on magnon topology, typically supplemented by the statements that they are anyway frozen out at low temperatures and/or negligible due to their $1/S$ smallness, where $S$ is the spin length. However, both statements do \emph{not} generally apply, rendering magnon-magnon interactions one of the most pressing issues in the field of magnon topology.

The few existing studies of how interactions between magnons influence topology have predominantly taken a pessimistic point of view by following the suspicion that interactions may render results obtained within the noninteracting limit obsolete. Chernyshev and Maksimov identified spontaneous magnon decays at zero temperature as the main obstacle \cite{Chernyshev2016}. For topological magnon insulators in kagome ferromagnets, they predicted a very large zero-field magnon broadening (damping) of the size of the topological gap. With the gap no longer well defined the notion of chiral in-gap states is jeopardized. Nonetheless, McClarty and Rau pointed out that since topological magnon gaps or band crossings necessarily occur at finite energy, the interaction-induced self-energy is non-Hermitian, opening up possibilities to explore non-Hermitian magnon topology \cite{McClarty2019}. An anisotropy of the magnon lifetime was identified as a crucial, indirect experimental signature of non-Hermitian topology. A direct detection of typical hallmarks of non-Hermitian topology, e.g., exceptional points, is still impeded by the global magnon linewidth broadening.

In principle, spontaneous magnon decays may be frozen out at high fields, which energetically separate the one-magnon from the two-magnon sector, rendering decays from the former into the latter kinematically impossible \cite{Rau2019}. Hence, the damping of magnons can be reduced in large fields, reinstating the notion of gaps and in-gap states. For example, field-polarized Kitaev magnets feature well-defined topological edge magnons despite interactions \cite{McClarty2018}.
Unfortunately, the field polarization trick is not always applicable. If the nontrivial magnon topology relies on noncollinearity as, for example, in kagome antiferromagnets \cite{Owerre2017a, Owerre2017c, Laurell2018, Mook2019}, large magnon damping is inevitable \cite{Chernyshev2009, Chernyshev2015, Chernyshev2015largeS}.
An exception to the rule are ferromagnetic skyrmion crystals that feature topologically nontrivial magnon bands at very low energy \cite{Molina2016, Garst2017, Diaz2019FMSkX}. Even for ultrasmall skyrmions built from but a few tens of magnetic moments, the energetically lowest nontrivial magnon band exhibits damping much smaller than the band gaps to its adjacent bands \cite{Mook2020SkyrmionDamping}. Hence, ferromagnetic skyrmion crystals are a unique platform to study magnonic topology and the validity of the noninteracting theory. Similar results may hold for ferromagnetic bimeron crystals \cite{Gobel2019bimeron} and antiferromagnetic skyrmion crystals \cite{Gobel2017}, a subclass of which \cite{Rosales2015} was shown to exhibit topological magnons at lowest possible energies \cite{Diaz2019}, thereby minimizing the kinematically allowed phase space for decays.

The above outlined state of the art suggests that magnon-magnon interactions need to be suppressed for Hermitian magnon topology to be appreciated. Herein, we explore particle-number nonconserving interactions beyond their detrimental lifetime broadening effect. Rather than analyzing if magnon topology is present \emph{in spite of interactions}, we look for nontrivial magnon topology \emph{because of interactions}.
Our main result is that interactions can break symmetries of the harmonic theory, causing topological phase transitions. Taking honeycomb ferromagnets as a contemporarily experimentally relevant example \cite{Yelon1971, Chen2018, Burch2018, Zhang2019wonder, Lu2019Curie, Kim2019, Zhang2019vdW, Xing2019, Yuan2020, Li2020vdW, Kashin2020, Liu2020honey, Sala2020arXiv}, we account for interactions within many-body perturbation theory to demonstrate how zero-temperature interactions cause a spontaneous mass gap of magnonic Dirac cones. We construct an effective theory to show that the interaction-induced mass gap is topologically nontrivial. Hence, a nontrivial Chern marker and winding number are found, signaling topologically protected chiral edge magnons in finite samples, a prediction that we explicitly verify numerically; the handedness of the chiral edge magnons is linked to the magnetization direction. 
This finding establishes the existence of nontrivial Hermitian magnon topology of genuinely quantum mechanical origin, i.e., due to spontaneous decays at order $1/S$, in sharp contrast to existing studies of Hermitian magnon topology, whose results could equally be arrived at by solving the linearized classical Landau-Lifshitz equation.

Our analysis reveals two unconventional mechanisms for topological magnon phase transitions from left to right-handed chirality (or vice versa): one field induced, the other temperature induced. We relate the field-induced transition to a topological transition in the two-particle decay contours, signaling a saddle point in the two-magnon continuum. The saddle point comes with a sign change in the interaction-induced Dirac mass term, which reverses Chern numbers. Hence, single-particle magnons can exhibit a rich topological phase diagram by virtue of their interplay with the two-magnon manifold. Moreover, we uncover the possibility of temperature-driven topological magnon phase transitions, brought about by thermally activated collision processes that compete with decays in that they cause Dirac masses of opposite sign. This result establishes temperature as an external control of topology, opening up new avenues in topological magnonics. Both types of topological phase transitions come with an experimental signature in transport experiments. Since nontrivial magnon topology has an impact on anomalous magnon-mediated transverse transport of spin and heat at finite temperatures \cite{Katsura2010, Matsumoto2011, Mook14a, Mook2016c}, the field or temperature-induced change in topology causes unconventional sign changes in spin Nernst and thermal Hall conductivities. 

Furthermore, we identify two working principles that allow for tailoring interactions in a way that their influence on magnon damping is suppressed, without compromising their effect on the band gap opening: (i) Using magnetic fields to tune Dirac cones close to the lower threshold of the two-magnon continuum. 
(ii) Externally \emph{enhancing} interactions because strong interactions---rather than completely wiping out the spectral weight of single-particle bands---expel the quasiparticle peaks from the continuum, reinstating long lifetimes. 
These two mechanisms, in principle, allow for arbitrarily long-lived topological magnons brought about by magnon-magnon interactions.

By means of exact diagonalization we extend our analysis to the ultimate quantum limit of $S=1/2$ and demonstrate the qualitative consistency between the semiclassical spin-wave theory and a full-quantum treatment. We conclude that, in principle, there can be a competition between noninteracting and interacting origins of topology, whose relative influence can be tuned by external fields. Hence, a priori, it is not sufficient to restrict a discussion of magnon topology to the noninteracting limit. We give an account of how interactions could influence the magnon topology and/or magnon-mediated anomalous transport of several experimentally available materials, which have previously been discussed in such contexts. 

Overall, our results demonstrate that magnon-magnon interactions should not be dismissed as being detrimental to magnon topology. Quite in contrast, they may be utilized to stabilize topology, a finding that gives rise to the paradigm of ``interacting topological magnonics'' and contributes to the fundamental understanding of interacting topological matter without particle number conservation, in which single and many-particle sectors cannot be considered separately.

The remainder of this paper is organized as follows. In Sec.~\ref{sec:Models}, we introduce two models of honeycomb-lattice ferromagnets---one ``achiral,'' the other ``chiral.'' These two models differ in their symmetries, with the achiral model featuring a Dirac cone stabilizing symmetry and the chiral model breaking this symmetry (Sec.~\ref{sec:Symmetry}). Within spin-wave theory, the symmetry-breaking part of the Hamiltonian is shown to cause particle-number nonconserving many-body interactions (Sec.~\ref{sec:SWT}), which we account for within lowest-order perturbation theory (Sec.~\ref{sec:ManyBodyPT}). These interactions are shown to gap out the Dirac cone of the chiral magnet (Sec.~\ref{sec:ResultsTopology}), introducing nontrivial topology (Sec.~\ref{sec:effectiveHam}) and chiral edge state (Sec.~\ref{sec:ResultsSlab}). The gap is demonstrated to depend on the magnetic field (Sec.~\ref{sec:ResultsTransitionField}), the strength of interactions (Sec.~\ref{sec:ResultsRevival}), and temperature (Sec.~\ref{sec:ResultsTransitionThermal}). A topological phase diagram is mapped out and the topological transitions are predicted to cause unconventional sign changes in transverse transport (Sec.~\ref{sec:ResultsTransport}). We complement the spin-wave analysis by exact diagonalization studies in Sec.~\ref{sec:ExactDiag}. After a discussion in Sec.~\ref{sec:Discussion}, we finally conclude in Sec.~\ref{sec:Conclusion}. Appendices \ref{sec:Vertices}-\ref{sec:AppendixSpin1} provide more detailed information on selected arguments.

% ====================================
%  MODELS
% ====================================
\section{Models: Achiral and chiral ferromagnets on the honeycomb lattice}
\label{sec:Models}

\begin{figure}
	\centering
	\includegraphics[scale=1]{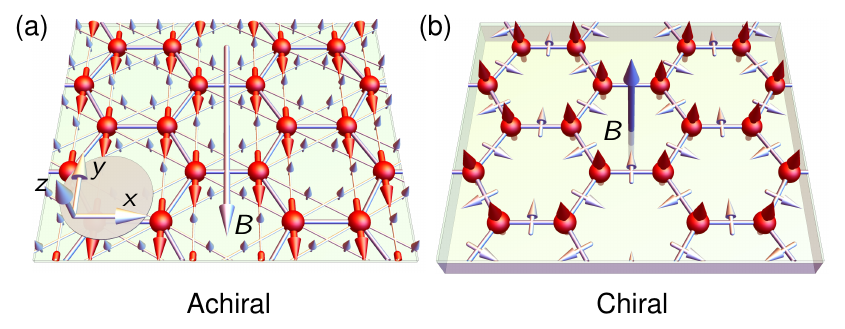}
	\caption{ 
		Honeycomb-lattice ferromagnets with Dzyaloshinskii-Moriya interaction (DMI) are minimal models to realize non-conserved interacting Dirac magnons. Red arrows depict field-polarized spin moments, small arrows at the bonds indicate DMI vectors, and the large arrow shows the direction of the magnetic field. (a) Achiral in-plane magnetized ferromagnet with second-nearest neighbor DMI. (b) Chiral ferromagnet with interfacial DMI due to structural inversion asymmetry, as indicated by a substrate layer. The texture is out-of-plane field polarized.
	}
	\label{fig:models}
\end{figure}

We consider both achiral (A) as well as chiral (C) two-dimensional ferromagnets on the honeycomb lattice, as depicted in Fig.~\ref{fig:models}. Their spin Hamiltonians read 
\begin{align}
	\hat{H}^\text{A/C} = \hat{H}_\mathrm{Z} + \hat{H}_\text{XC} + \hat{H}_\text{DMI}^\text{A/C},
	\label{eq:full-ham}
\end{align}
where the Zeeman energy
\begin{align}
	\hat{H}_\mathrm{Z} = - \sum_{\vec{r}} \vec{B} \cdot \hat{\vec{S}}_{\vec{r}},
\end{align}
accounts for a magnetic field $\vec{B}$ that acts on all spin operators $\hat{\vec{S}}_{\vec{r}}$ ($\vec{r}$ points to a lattice site). In both cases, we consider 
\begin{align}
	\hat{H}_\text{XC} = -\frac{J}{2} \sum_{\langle \vec{r}, \vec{r}' \rangle} \hat{\vec{S}}_{\vec{r}} \cdot \hat{\vec{S}}_{\vec{r}'},
\end{align}
with positive Heisenberg exchange $J$ between nearest neighbors stabilizing a ferromagnetic phase. 
The two models only differ in their antisymmetric exchange 
\begin{subequations}
\begin{align}
    \hat{H}^\text{A}_\text{DMI} 
    &= 
    \frac{D_z}{2} \sum_{\langle \langle \vec{r}, \vec{r}' \rangle \rangle}  \nu_{\vec{r},\vec{r}'} \vec{z} \cdot \left( \hat{\vec{S}}_{\vec{r}} \times \hat{\vec{S}}_{\vec{r}'} \right),
    \label{eq:Ham_achiral}
    \\
    \hat{H}^\text{C}_\text{DMI} 
    &= 
    \frac{D}{2} \sum_{\langle \vec{r}, \vec{r}' \rangle} \vec{d}_{\vec{r},\vec{r}'} \cdot \left( \hat{\vec{S}}_{\vec{r}} \times \hat{\vec{S}}_{\vec{r}'} \right).
    \label{eq:Ham_chiral}
\end{align}
\end{subequations}
In Eq.~\eqref{eq:Ham_achiral}, the out-of-plane relativistic Dzyaloshinskii-Moriya interaction \cite{Dzyaloshinsky58, Moriya60} (DMI) $D_z$ acts between second-nearest neighbors, with $\nu_{\vec{r},\vec{r}'} = \pm 1$ chosen negative (positive) for (counter-)clockwise circulation around a hexagon [cf.~Fig.~\ref{fig:models}(a); clockwise circulation applies]. This type of DMI does not compromise the classical ferromagnetic ground state; hence the name ``achiral magnet.'' To select a particular axis of magnetization, the magnetic field is applied within the honeycomb plane; in Fig.~\ref{fig:models}(a), we take $\vec{B} \parallel - \vec{y}$.

In contrast, in Eq.~\eqref{eq:Ham_chiral}, the interfacial nearest-neighbor DMI $D$, with $\vec{d}_{ij} = \vec{z} \times (\vec{r}_j - \vec{r}_i)/|\vec{r}_j - \vec{r}_i|$  [cf.~Fig.~\ref{fig:models}(b)], favors noncollinear ordering, rendering the magnetic ground state a N\'{e}el spin spiral with a fixed chirality; hence the name ``chiral magnet.'' Applying $\vec{B} \parallel \vec{z}$ out of the plane, the classical ground state first changes into a skyrmion crystal \cite{Muhlbauer2009}, and then into the field-polarized phase, which is the case approximately for $|\vec{B}| \gtrsim 0.8 D^2/J$ \cite{Han2010}. In the remainder of this paper, we exclusively focus on the limit of field polarization.

% ====================================
%  SYMMETRY
% ====================================
\subsection{Symmetry analysis}
\label{sec:Symmetry}
For discussing symmetries, we have to consider both the Hamiltonian and its magnetically ordered classical ground state. We drop the hat symbol to indicate ground state spin directions ($\hat{\vec{S}}_{\vec{r}} \to \vec{S}_{\vec{r}}$).

Spectral Dirac points on the honeycomb lattice are stable in the simultaneous presence of inversion symmetry ($\mathcal{P}$; also called sublattice or parity symmetry) and time-reversal symmetry ($\mathcal{T}$). Although ferromagnetism violates actual $\mathcal{T}$ symmetry (spin flip), there can be an effective time-reversal symmetry $\mathcal{T}' = \mathcal{R}(\vec{n},\pi) \mathcal{T}$, composed from $\mathcal{T}$ and a rotation $\mathcal{R}(\vec{n},\pi)$ by $\pi$ in spin space about an axis $\vec{n}$ normal to the magnetization $\vec{M} \equiv \frac{1}{N} \sum_{\vec{r}}^N \vec{S}_{\vec{r}}$ to map the texture back onto itself ($N$ is the total number of spins). If this rotation leaves all terms of the Hamiltonian invariant, $\mathcal{T}'$ is a good symmetry of the magnet.

\paragraph*{Achiral magnet.} For $\vec{M}$ within the honeycomb plane, as in Fig.~\ref{fig:models}(a), we find that $\mathcal{T}' = \mathcal{R}(\vec{z},\pi) \mathcal{T}$ is a symmetry of the system. After the action of $\mathcal{T}$ has flipped the spins, a rotation $\mathcal{R}(\vec{z},\pi)$ by $\pi$ about the $z$ axis maps the texture back onto itself and leaves the $D_z$ terms invariant. To see so, recall that
\begin{align}
	\mathcal{R}(\vec{z},\pi) \vec{S}_{\vec{r}} 
	= 
	( -S^x_{\vec{r}}, -S^y_{\vec{r}}, S^z_{\vec{r}} ), \label{eq:rotation}
\end{align}
which has no effect on
$
	D_z ( S^x_{\vec{r}} S^y_{\vec{r}'} - S^y_{\vec{r}} S^x_{\vec{r}'} )
$.

Moreover, although DMI relies on local inversion asymmetry at a bond's midpoint \cite{Moriya60}, the achiral magnet is $\mathcal{P}$ invariant, with the center of inversion being located at a hexagon's center of mass. (Recall that both the spins as well as the DMI vectors are axial vectors.)
Hence, both stabilizing symmetries of the Dirac cone are present.

\paragraph*{Chiral magnet.} There is no possibility to construct a $\mathcal{T}'$ symmetry for the chiral magnet in Fig.~\ref{fig:models}(b) because there is no in-plane axis about which the spins can be rotated back without changing the DMI terms. This lack of a rotation axis traces back to the DMI vectors spanning a plane. Hence, the chiral magnet breaks $\mathcal{T}'$ symmetry.

Surprisingly, in spite of broken interfacial inversion symmetry, there is an effective parity symmetry $\mathcal{P}' = \mathcal{R}(\vec{z},\pi) \mathcal{P}$. After the action of $\mathcal{P}$ the substrate is on top of the magnet and, hence, the DMI vectors are opposite; it amounts to the mapping $D \to -D$. By a $\mathcal{R}(\vec{z},\pi)$ rotation in spin space, we compensate for the minus sign, because the DMI terms always connect $S^z_{\vec{r}}$ with an in-plane spin component, the latter of which acquires a minus sign upon rotation [cf.~Eq.~\eqref{eq:rotation}]. Hence, the parity-breaking effect of interfacial DMI can be ``rotated away.''

\begin{table}
	\centering
	\caption{Presence or absence of Dirac-cone stabilizing symmetries in the two model systems depicted in Fig.~\ref{fig:models}.}
	\begin{tabular}{lcc}
		\toprule
		                       & Achiral [Fig.~\ref{fig:models}(a)] & Chiral [Fig.~\ref{fig:models}(b)] \\
		\hline
		Time-reversal symmetry & yes           & no \\
		Parity symmetry        & yes           & yes \\
		\botrule
	\end{tabular}
	\label{tab:symmetries}
\end{table}
 
As summarized in Tab.~\ref{tab:symmetries}, the achiral magnet obeys both symmetries, but the chiral magnet is only parity symmetric. Hence, stable Dirac points are expected for achiral magnets but a mass gap for chiral magnets.
Although these symmetry arguments appear to be straightforward, the situation is more intricate, as we elobarate on in the following section.

% ====================================
%  SPIN-WAVE THEORY
% ====================================
\subsection{Spin-Wave theory}
\label{sec:SWT}
Relying on a magnetically ordered classical ground state, we proceed with a Holstein-Primakoff (HP) transformation \cite{Holstein1940}
\begin{align}
    \hat{\vec{S}}_{\vec{r}} 
    = 
    \sqrt{S} \left( \hat{f}_{\vec{r}} \hat{a}_{\vec{r}} \vec{e}_{\vec{r}}^{-} + \hat{a}_{\vec{r}}^\dagger \hat{f}_{\vec{r}} \vec{e}_{\vec{r}}^{+} \right) 
    + \left( S - \hat{a}_{\vec{r}}^\dagger \hat{a}_{\vec{r}}\right) \vec{e}_{\vec{r}}^{z},
\end{align}
where $S$ is the spin length. The axes of the local reference frame read $\vec{e}_{\vec{r}}^\pm = ( 1, 0, \pm\mathrm{i} )/\sqrt{2}$, and $\vec{e}_{\vec{r}}^{z} = (0,-1,0)$ for the achiral magnet and $\vec{e}_{\vec{r}}^\pm = ( 1, \pm\mathrm{i}, 0 )/\sqrt{2}$, and $\vec{e}_{\vec{r}}^{z} = (0,0,1)$ for the chiral magnet; $\mathrm{i}^2 = -1$.
The bosonic operators obey the usual commutation relation 
$
	[\hat{a}_{\vec{r}},\hat{a}^\dagger_{\vec{r}'}] = \delta_{\vec{r}, \vec{r}'} 
$. 
A Taylor expansion of
\begin{align}
    \hat{f}_{\vec{r}} = 
    \left( 1- \frac{\hat{a}_{\vec{r}}^\dagger \hat{a}_{\vec{r}}}{2S} \right)^{1/2}
    =
    1 - \frac{1}{2} \frac{\hat{a}_{\vec{r}}^\dagger \hat{a}_{\vec{r}}}{2S}
    - \ldots 
    \label{eq:f}
\end{align}
effectively expands the Hamiltonian as $\hat{H} = \sum_{p=0}^\infty \hat{H}_p$, where $p$ denotes the number of bosonic operators. Formally, this is an expansion in $1/\sqrt{S}$, with $\hat{H}_p \propto O[(S^{-1})^{p/2 - 2}]$. Up to quartic order, Hamiltonian \eqref{eq:full-ham} reads 
\begin{align}
    \hat{H}^\text{A/C} \approx E_0 + \hat{H}_2 + \hat{H}_3^\text{A/C} + \hat{H}_4.
\end{align}
Importantly, the two magnets differ only in their DMI-induced \emph{number non-conserving} three-particle Hamiltonian $\hat{H}_3^\text{A/C}$. The remaining terms, i.e., the ground state energy $E_0$, the bilinear Hamiltonian $\hat{H}_2$, and the number conserving four-particle Hamiltonian $\hat{H}_4$, derive from the exchange energy, which is identical in both models.

\begin{figure}
	\centering
	\includegraphics[scale=1]{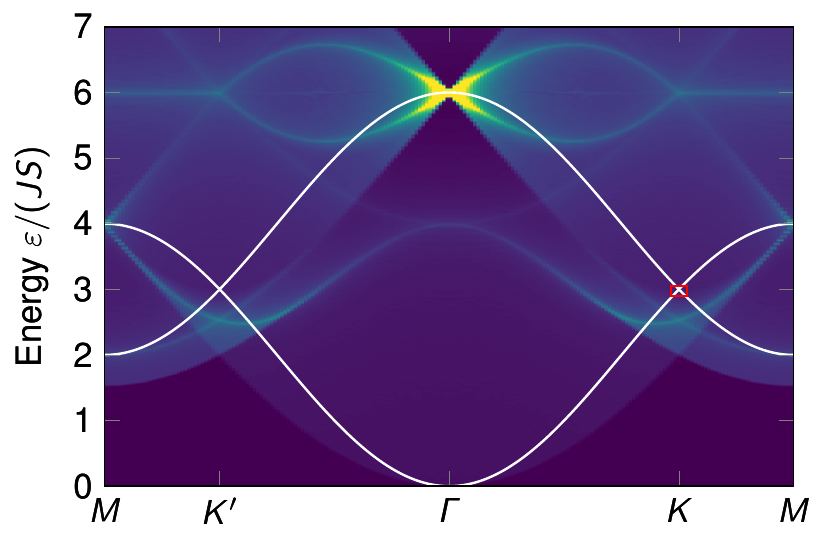}
	\caption{ 
		Single and two-magnon sectors in honeycomb ferromagnets along high-symmetry paths of the Brillouin zone. At $B=0$, the harmonic single-particle energies [white lines; cf.~Eq.~\eqref{eq:single-particle-energy}] overlap with the two-magnon density of states [color plot; cf.~Eq.~\eqref{eq:twomagnonDOS}]. Dark blue/yellow color indicates zero/maximal two-magnon density of states. The red rectangle around the Dirac cone at the $K$ point indicates the momentum and energy window shown in Fig.~\ref{fig:tomographic-spectral-function}.
		For increasing fields, the single-particle energies are shifted upwards in energy by $B$, while the two-magnon continuum is shifted by $2B$. Hence, for large enough fields, the continuum is shifted past the single-particle energies. In particular, the Dirac cones at the $K$ and $K'$ points leave the continuum at $B/(JS) = 1$.
	}
	\label{fig:DOS}
\end{figure}

\paragraph*{Harmonic Theory.}
Linear spin-wave theory amounts to diagonalizing the Fourier transformed quadratic Hamiltonian 
$
    \hat{H}_2 = \sum_{\vec{k}} \hat{\vec{A}}^\dagger_{\vec{k}} \cdot \mathcal{H}_{\vec{k}} \cdot \hat{\vec{A}}_{\vec{k}}
$, 
with 
$
    \hat{\vec{A}}_{\vec{k}}^\dagger = (\hat{a}^\dagger_{\vec{k},1}, \hat{a}^\dagger_{\vec{k},2})
$ 
and
\begin{align}
    \mathcal{H}_{\vec{k}} = 
    \begin{pmatrix}
		3JS + B & -JS\gamma_{\vec{k}} \\
		-JS\gamma_{-\vec{k}} & 3JS + B
	\end{pmatrix},
	\quad
	\gamma_{\vec{k}} = \sum_{i=1}^3 \mathrm{e}^{\mathrm{i} \vec{k} \cdot \vec{\delta}_i},
	\label{eq:HamiltonMarix}
\end{align}
by a unitary transformation to the normal modes
$
    \hat{\vec{B}}_{\vec{k}} = \mathcal{U}^\dagger_{\vec{k}} \hat{\vec{A}}_{\vec{k}}
$,
where
$
    \hat{\vec{B}}_{\vec{k}}^\dagger = (\hat{b}^\dagger_{\vec{k},-}, \hat{b}^\dagger_{\vec{k},+})
$.
The $\hat{a}_{\vec{k},i}$'s are the Fourier transformed bosonic annihilators of a spin deviation (or HP boson) at the $i$th basis site. Similarly, the $\hat{b}_{\vec{k},j}$'s are the annihilators of a magnon in the $j$th band ($j=\pm$).
The off-diagonal $\gamma_{\vec{k}}$ contains the nearest neighbor bonds
\begin{subequations}
\begin{align}
	\vec{\delta}_1 &= (1,0), \label{eq:delta1}\\
	\vec{\delta}_2 &= (-1/2,\sqrt{3}/2), \\
	\vec{\delta}_3 &= (-1/2,-\sqrt{3}/2). \label{eq:delta3}
\end{align}
\end{subequations}
The distance between nearest neighbors is set to one.
Using
\begin{align}
	\mathcal{U}_{\vec{k}} = \frac{1}{\sqrt{2}}
	\begin{pmatrix}
		\mathrm{e}^{\mathrm{i} \xi_{\vec{k}}/2} & \mathrm{e}^{\mathrm{i} \xi_{\vec{k}}/2} \\
		\mathrm{e}^{-\mathrm{i} \xi_{\vec{k}}/2} & -\mathrm{e}^{-\mathrm{i} \xi_{\vec{k}}/2} 
	\end{pmatrix},
	\quad
	\xi_{\vec{k}} = \mathrm{arg} ( \gamma_{\vec{k}} ),
	\label{eq:eigenvectors}
\end{align}  
we find that 
$
    \mathcal{E}_{\vec{k}} = \mathcal{U}^\dagger_{\vec{k}} \mathcal{H}_{\vec{k}} \mathcal{U}_{\vec{k}} = \mathrm{diag}(\varepsilon_{\vec{k},-},\varepsilon_{\vec{k},+})
$ 
contains the single-particle magnon energies
\begin{align}
	\varepsilon_{\vec{k},\pm} = JS\left( 3 \pm |\gamma_{\vec{k}}| \right) + B. \label{eq:single-particle-energy}
\end{align}
(Note again that $B \gtrsim 0.8 D^2/J$ must hold for the chiral magnet to become field polarized \cite{Han2010}, i.e., for the ferromagnetic phase to be the ground state.)
Hence, the spectrum of the free theory is identical to that of ``spinless'' electrons on the honeycomb lattice. The magnonic collective excitations, which come in two flavors, exhibit a graphene-like band structure, as depicted in Fig.~\ref{fig:DOS} by white lines. The two bands touch linearly at the Dirac point energy 
\begin{align}
	\varepsilon_\text{D} = 3JS+B
	\label{eq:DiracEnergy}
\end{align}
at the corners ($K$ and $K'$ points) of the hexagonal unit cell, where $|\gamma_{\vec{k}}| = 0$. From hereinafter, we measure the excitation energies relative to $\varepsilon_\text{D}$, such that the harmonic Dirac cone appears at ``zero energy.''

The Dirac cones have zero mass (no gap). Since the harmonic theory of a ferromagnet without quantum fluctuations coincides with the linearized equation of motion of classical spin vectors (Landau-Lifshitz equation without damping), the Dirac magnons are also said to have \textit{zero classical mass}.

\paragraph*{Anharmonic Theory.}
Note that the DMI did not enter the harmonic theory. This is because DMI vectors orthogonal to the magnetization direction connect longitudinal with transverse spin components (e.g., $\hat{S}_i^z \hat{S}_j^{x,y}$ in a local reference frame, where $z$ is chosen along the magnetization direction), which contribute to the $\hat{H}_p$'s with odd $p$. The contributions to $\hat{H}_1$ cancel, such that the lowest-order appearance of the DMI is in $\hat{H}_3$, which comprises magnon decay (one magnon decays into two) and coalescence (two magnons coalesce into one). This particle number nonconserving interaction is in line with the DMI-induced nonconservation of spin, i.e., $[ \sum_i \hat{\vec{S}}_i \cdot \vec{m}, \hat{H}] \propto D$ (or $\propto D_z$), with $\vec{m} = \vec{M}/M$ being the magnetization direction. Thus, the spin-orbit coupling-induced DMI connects the magnons to the lattice bath, into (from) which angular momentum can be dumped (drawn) during a magnon coalescence (decay).

The cubic Hamiltonian reads
\begin{align}
    \hat{H}_3^\text{A/C} = 
    \frac{1}{2 \sqrt{N}} \sum_{\lambda, \mu, \nu = 1}^2 \sum_{\vec{k}, \vec{q}, \vec{p}}^{\vec{p}=\vec{k}+\vec{q}} \left[
        \left( V^{\lambda\mu \leftarrow \nu}_{\vec{k}, \vec{q} \leftarrow \vec{p}} \right)^\text{A/C}
        \hat{a}^\dagger_{\vec{k}, \lambda}  \hat{a}^\dagger_{\vec{q}, \mu}  \hat{a}_{\vec{p}, \nu} 
        +
        \mathrm{H.c.} \right]
        \label{eq:H3HP}
\end{align}
and accounts for an incoming HP boson $\hat{a}_{\vec{p}, \nu}$ with momentum $\vec{p}$ being destroyed but two bosons $\hat{a}^\dagger_{\vec{k}, \lambda}$ and $\hat{a}^\dagger_{\vec{q}, \mu}$ being created. This process has to obey momentum conservation, $\vec{p} = \vec{k}+\vec{q}$, modulo reciprocal lattice vectors.
The Hermitian conjugate (H.c.) part in Eq.~\eqref{eq:H3HP} describes the opposite process with two incoming and one outgoing boson. The HP interaction vertices $\left( V^{\lambda\mu \leftarrow \nu}_{\vec{k}, \vec{q} \leftarrow \vec{p}} \right)^\text{A/C}$ are given in Appendix \ref{sec:Vertices}. Importantly, they are linear in DMI $D_z$ (achiral magnet) or $D$ (chiral magnet) and independent of the exchange energy $J$.

Eventually, we work in the eigenbasis, such that the cubic Hamiltonian must be expressed in terms of normal modes,
\begin{align}
    \hat{H}_3^\text{A/C} = 
    \frac{1}{2 \sqrt{N}} \sum_{\lambda, \mu, \nu=\pm} \sum_{\vec{k}, \vec{q}, \vec{p}}^{\vec{p}=\vec{k}+\vec{q}} \left[
        \left( \mathcal{V}^{\lambda \mu \leftarrow \nu}_{\vec{k}, \vec{q} \leftarrow \vec{p}} \right)^\text{A/C}
        \hat{b}^\dagger_{\vec{k}, \lambda}  \hat{b}^\dagger_{\vec{q}, \mu}  \hat{b}_{\vec{p}, \nu} 
        +
        \mathrm{H.c.} \right]
        ,
        \label{eq:H3}
\end{align}
where the three-magnon vertices
\begin{align}
    \left( \mathcal{V}^{\lambda\mu \leftarrow \nu}_{\vec{k}, \vec{q} \leftarrow \vec{p}} \right)^\text{A/C}
    =
    \sum_{l,m,n = 1}^2
    \left( V^{l m \leftarrow n}_{\vec{k}, \vec{q} \leftarrow \vec{p}}  \right)^\text{A/C}
    \mathcal{U}^\ast_{\vec{k},l\lambda} \mathcal{U}^\ast_{\vec{q}, m\mu} \mathcal{U}_{\vec{p}, n\nu}
\end{align}
are constructed from the HP vertices and the eigenvectors given in Eq.~\eqref{eq:eigenvectors}.

At higher-order spin-wave theory, we encounter the number conserving four-magnon interactions ($\hat{H}_4$), which derive from the exchange energy. (The explicit expression of $\hat{H}_4$ can be found in Ref.~\onlinecite{Pershoguba2018}.) Apart from being frozen out at zero temperature anyway, $\hat{H}_4$ shares the symmetries of $\hat{H}_2$ and, hence, does not play a significant role. Section \ref{sec:ManyBodyPT} gives more detailed arguments.

We close with noting that we are faced with the theoretically intriguing situation that particular magnetic interactions (here: DMI) between spins appear exclusively as number non-conserving interactions between excitations. Since DMI is the crucial ingredient to break the effective time-reversal symmetry of the chiral magnet (cf.~Sec.~\ref{sec:Symmetry}), DMI-derived interactions break accidental symmetries of the harmonic theory and a highly nontrivial renormalization of the Dirac cones is expected.
Since it does not take any thermal excitation for a magnon to \emph{spontaneously} decay, the genuinely quantum-mechanical effects of $\hat{H}^\text{A/C}_3$ persist down to zero temperature, which is the limit we will explore first. Any gap of the Dirac cone is hence associated with a \textit{spontaneous quantum mass}.

% ====================================
%  Many-body perturbation theory
% ====================================
\subsection{Many-body perturbation theory}
\label{sec:ManyBodyPT}

To account for the effects of $\hat{H}^\text{A/C}_3$ and $\hat{H}_4$, we invoke many-body perturbation theory for the single-particle Green's function, from which we ultimately extract the renormalized magnon spectrum in terms of the spectral function. The latter is related to the dynamical structure factor (i.e., the spin-spin correlation function) probed in neutron scattering experiments \cite{Lovesey1977}. 

Up to order $1/S$, the interacting (Matsubara) one-magnon Green's function is given by \cite{Rastelli2011}
\begin{align}
    &\mathcal{G}_{\vec{k},\alpha \beta}(\tau) 
    \approx \mathcal{G}^{(0)}_{\vec{k},\alpha \beta}(\tau) 
    +\int_0^{\beta} \mathrm{d} \tau_1  \left\langle 
    \mathcal{T}_\tau \hat{H}_4(\tau_1) \hat{b}_{\vec{k},\alpha}(\tau) \hat{b}^\dagger_{\vec{k},\beta}
    \right\rangle_{(0)}^{\mathrm{con}}
    \nonumber \\
    &\quad - \frac{1}{2!} \int_0^{\beta} \mathrm{d} \tau_1 \int_0^{ \beta} \mathrm{d} \tau_2 \left\langle
    \mathcal{T}_\tau \hat{H}_3(\tau_1) \hat{H}_3(\tau_2) \hat{b}_{\vec{k},\alpha}(\tau) \hat{b}^\dagger_{\vec{k},\beta}
    \right\rangle_{(0)}^{\mathrm{con}},
    \label{eq:Green-perturb}
\end{align}
where 
$
	\mathcal{G}^{(0)}_{\vec{k},\alpha \beta}(\tau) = - \langle \mathcal{T}_\tau  \hat{b}_{\vec{k},\alpha}(\tau) \hat{b}^\dagger_{\vec{k},\beta} \rangle_{(0)}
$ 
is the noninteracting Green's function, $\beta = (k_\text{B} T)^{-1}$ is inverse temperature ($k_\text{B}$ being Boltzmann's constant), $\mathcal{T}_\tau$ orders imaginary times $\tau_i$, and averages $\langle \cdot \rangle_{(0)}^{\mathrm{con}}$ are taken with respect to $\hat{H}_2$ over connected (`con') diagrams.

\begin{figure}
	\centering
	\includegraphics[scale=1]{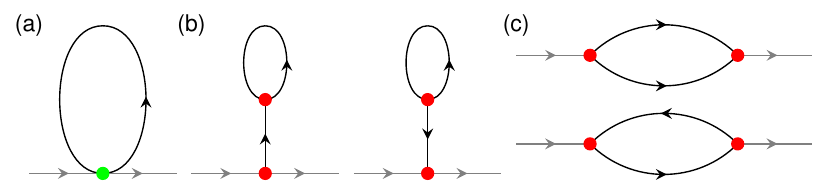}
	\caption{ 
		Feynman diagrams contributing to $1/S$ many-body corrections in anharmonic spin-wave theory.
		(a) Hartree diagram with the number-conserving four-magnon vertex indicated by a green circle. (b) Upwards and downwards tadpole diagrams, with the three-magnon vertices indicated by red circles. (c) Forward (top) and circle (bottom) bubble diagrams. At zero temperature, only the forward bubble diagram is contributing to the self-energy.
	}
	\label{fig:feynman}
\end{figure}

After applying Wick's theorem, the first-order perturbation proportional to $\hat{H}_4$ yields self-energies associated with the Hartree diagram in Fig.~\ref{fig:feynman}(a). For the second-order perturbation ($\hat{H}_3$), one finds $4$ ``forward bubble'' diagrams [upper diagram in Fig.~\ref{fig:feynman}(c)], and $8$ each of ``upwards tadpole'' [left diagram in Fig.~\ref{fig:feynman}(b)], ``downwards tadpole'' [right diagram in Fig.~\ref{fig:feynman}(b)], and ``circle bubble'' [lower diagram in Fig.~\ref{fig:feynman}(c)].

\paragraph*{Hartree diagram.}
The Hartree contribution in Fig.~\ref{fig:feynman}(a) is most conveniently evaluated before the diagonalization procedure, as demonstrated by Pershoguba \textit{et al.}~\cite{Pershoguba2018}. Extending their analysis to finite fields, we may approximate
\begin{align}
	\varSigma^\text{Hartree}_{\vec{k}}(T)
	\approx
	- \frac{\lambda}{\beta^2} \mathcal{H}_{\vec{k}}
\end{align}
(given in HP basis $\hat{\vec{A}}_{\vec{k}}$) at low temperatures, $\beta (SJ+B) \gg 1$. The constant
\begin{align}
	\lambda = \frac{\mathrm{Li}_2 \left( \mathrm{e}^{-\beta B} \right)}{4 \sqrt{3} \pi J^2 S^3}
	\label{eq:lambda}
\end{align}
contains a thermal-activation-like factor (with the dilogarithm $\text{Li}_2$), which accounts for field-freezing of four-magnon interactions.\footnote{In the limit $\beta B \to 0$, one obtains $\mathrm{Li}_2 ( \mathrm{e}^{-\beta B} ) \to \pi^2/6$ and $\lambda \to \pi/(24 \sqrt{3} \pi J^2 S^3)$, agreeing with the zero-field result of Ref.~\onlinecite{Pershoguba2018}.} Hence, without DMI, the renormalized magnon energies read
\begin{align}
	\tilde{\varepsilon}_{\vec{k,\pm}} = \left( 1-\frac{\lambda}{\beta^2} \right) \varepsilon_{\vec{k,\pm}}.
\end{align}
There are neither interaction-induced band gaps nor magnon lifetimes. 
The spectrum is merely compressed which reduces the Dirac velocity. Note also the $1/S^3$ dependence of $\lambda$ in Eq.~\eqref{eq:lambda},  which reveals that the low-temperature influence of Hartree contributions is even weaker than the nominal order $1/S$.
Hence, we disregard the Hartree contributions in the remainder of our study.

\paragraph*{Tadpole diagrams.}
The self-energy of the two tadpole diagrams in Fig.~\ref{fig:feynman}(b) reads
\begin{align}
	\varSigma_{\vec{k}}^{\text{tad},\alpha \beta}(T) 
    = 
      \frac{1}{N} \sum_{\vec{q} \in \text{BZ}} \sum_{j, j'=\pm} 
      &\left(
      	\mathcal{V}^{\alpha \leftarrow \beta j }_{ \vec{k} \leftarrow \vec{k}, \vec{0}}
        \mathcal{V}^{j j' \leftarrow j'}_{\vec{0}, \vec{q} \leftarrow \vec{q}}
        +
        \mathcal{V}^{j' \leftarrow j j'}_{\vec{q}\leftarrow \vec{0}, \vec{q}}
        \mathcal{V}^{\alpha j \leftarrow \beta }_{ \vec{k}, \vec{0} \leftarrow \vec{k}}
      \right)
       \nonumber \\
	  &\quad \times\frac{\rho(\varepsilon_{\vec{q},j'},T)}{\varepsilon_{\vec{0},j}}      
    \label{eq:Sigma-tadpole}
\end{align}
(given in eigenmode basis $\hat{\vec{B}}_{\vec{k}}$). The summation is over momenta $\vec{q}$ in the Brillouin zone (BZ) and the indices $\alpha,\beta=\pm$ label a self-energy matrix element.
Due to the Bose factor, this self-energy is zero at $T=0$ and, hence, a priori plays an insignificant role. Moreover, our numerical evaluation of Eq.~\eqref{eq:Sigma-tadpole} revealed that $\varSigma_{\vec{k}}^{\text{tad},\alpha \beta}(T)$ is zero at all temperatures, rendering tadpole diagrams irrelevant.

\paragraph*{Bubble diagrams.}
We are thus left with ``bubbles'' [Fig.~\ref{fig:feynman}(c)], whose self-energy (given in eigenmode basis $\hat{\vec{B}}_{\vec{k}}$)
\begin{align}
    \varSigma_{\vec{k}}^{\alpha \beta}(\varepsilon,T) 
    = 
      \frac{1}{N} \sum_{\vec{q} \in \text{BZ}} \sum_{j, j'=\pm} 
      &\left(
      \frac{1}{2} 
	  \frac{\mathcal{V}^{\alpha \leftarrow j j' }_{ \vec{k} \leftarrow \vec{q}, \vec{k}-\vec{q} }
        \mathcal{V}^{j j' \leftarrow \beta}_{\vec{q}, \vec{k}-\vec{q} \leftarrow \vec{k}} 
        }{\varepsilon + \mathrm{i} 0^+ - \varepsilon_{\vec{q},j} - \varepsilon_{\vec{k}-\vec{q},j'} }
        \right.
        \nonumber \\
      & \times \left[ \rho(\varepsilon_{\vec{q},j},T) + \rho(\varepsilon_{\vec{k}-\vec{q},j'},T) + 1 \right]
      \nonumber \\
    & + 
      \frac{
      	\mathcal{V}^{ j' \leftarrow \beta j }_{ \vec{k}+\vec{q} \leftarrow \vec{k}, \vec{q}} \mathcal{V}^{\alpha j \leftarrow j'}_{ \vec{k}, \vec{q} \leftarrow \vec{k}+\vec{q} }  
      	}{ 
      	\varepsilon + \mathrm{i} 0^+ + \varepsilon_{\vec{q},j} - \varepsilon_{\vec{k}+\vec{q},j'} }
      	\nonumber \\
      &\times \left.
      \left[
      	\rho(\varepsilon_{\vec{q},j},T) - \rho(\varepsilon_{\vec{k}+\vec{q},j'},T)
      \right]
      \right)
    \label{eq:Sigma-bubble}
\end{align}
has two contributions.
The first term of Eq.~\eqref{eq:Sigma-bubble} derives from the ``forward bubbles'' and is called ``decay term'' and the second one from the ``circle bubbles'' and is coined ``collision term.''\footnote{The numerical factor of $1/2$ in front of the decay term in Eq.~\eqref{eq:Sigma-bubble} derives from the observation that the Wick expansion only yields 4 ``forward bubbles'' but 8 ``circle bubbles.'' These numbers get multiplied by $(1/2)^3$, where one $1/2$ arises from the second-order perturbative expansion in Eq.~\eqref{eq:Green-perturb} and the other two from the definition of $\hat{H}_3$ in Eq.~\eqref{eq:H3}.}

% ====================================
%  Band gap at T=0
% ====================================
\section{Results}
\subsection{Interaction-induced gap opening at zero temperature: Spontaneous quantum mass}
\label{sec:ResultsGap}
To analyze the many-body renormalized magnon spectrum, we calculate the spectral function
\begin{align}
	A_{\vec{k}} (\varepsilon) = - \frac{1}{\pi} \mathrm{Im} \{ \mathrm{Tr} [ \mathcal{G}_{\vec{k}}(\varepsilon) ] \}
\end{align}
from the retarded Green's function matrix 
\begin{align}
	\mathcal{G}_{\vec{k}}(\varepsilon) = \left[ \varepsilon + \mathrm{i}0^+ - \mathcal{E}_{\vec{k}} - \varSigma_{\vec{k}}(\varepsilon) \right]^{-1}.
\end{align}
At $T=0$, only the spontaneous contribution from the $+1$ summand in Eq.~\eqref{eq:Sigma-bubble} remains and the self-energy reads
\begin{align}
    \varSigma_{\vec{k}}^{\alpha \beta}(\varepsilon) 
    = 
      \frac{1}{2N} \sum_{\vec{q} \in \text{BZ}} \sum_{j, j' = \pm} 
	  \frac{\mathcal{V}^{\alpha \leftarrow j j' }_{ \vec{k} \leftarrow \vec{q}, \vec{k}-\vec{q} }
        \mathcal{V}^{j j' \leftarrow \beta}_{\vec{q}, \vec{k}-\vec{q} \leftarrow \vec{k}} 
        }{\varepsilon + \mathrm{i} 0^+ - \varepsilon_{\vec{q},j} - \varepsilon_{\vec{k}-\vec{q},j'} }.
        \label{eq:self-energy-at-zeroT}
\end{align}
Hereinafter, we take $S=1$ if not stated otherwise.

\begin{figure}
	\centering
	\includegraphics[scale=1]{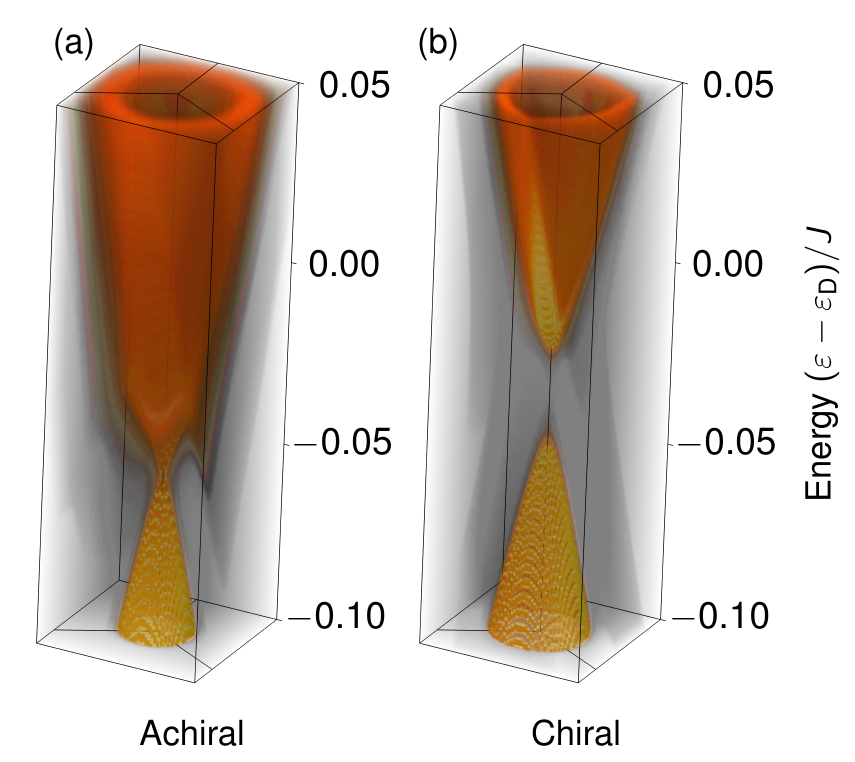}
	\caption{ 
		Tomographic cut of the interaction-renormalized magnon Dirac cones in (a) the achiral and (b) chiral magnet. Black transparent/orange opaque color indicates zero/maximal spectral function $A_{\vec{k}} (\varepsilon)$. The momentum and energy window shown here coincides with that marked by a red rectangle in Fig.~\ref{fig:DOS}. Parameters read $S=1$, $B/J=1$, and (a) $D_z/J=0.15$, (b) $D/J=0.15$.
	}
	\label{fig:tomographic-spectral-function}
\end{figure}

We evaluate $A_{\vec{k}} (\varepsilon)$ in the vicinity of the harmonic Dirac cones (cf.~red window in Fig.~\ref{fig:DOS}) both for the achiral and chiral magnet at $B/(JS) = 1$; results are shown in Fig.~\ref{fig:tomographic-spectral-function}(a) and (b), respectively. 
We see that the Dirac cone of the achiral magnet, although being renormalized both in energy as well as ``sharpness'' (lifetime broadening), is stable. In contrast, signatures of a spectral gap opening are witnessed for the chiral magnet, suggesting that the Dirac magnons acquired a spontaneous quantum mass. These findings are in accord with the symmetry analysis in Sec.~\ref{sec:Symmetry}.

\begin{figure}
	\centering
	\includegraphics[scale=1]{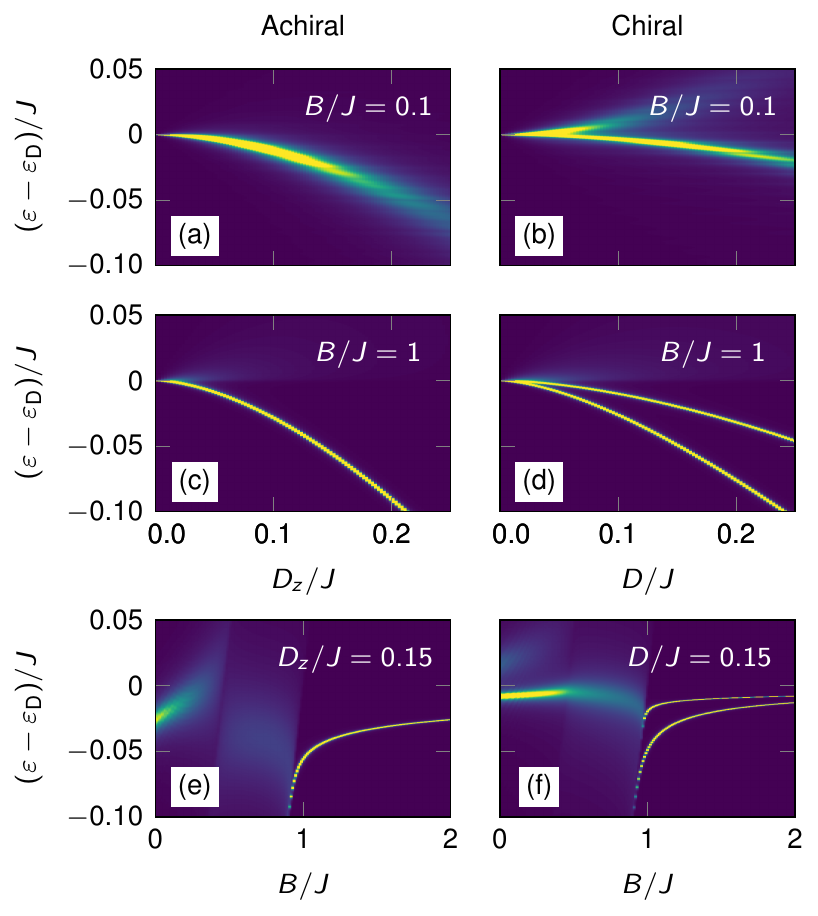}
	\caption{ 
		Magnonic spectral function $A_{\vec{K}}(\varepsilon)$ at the $K$ point in dependence on DMI $D$ and $D_z$ and magnetic field $B$ for the achiral (left) and chiral (right) magnet, respectively. Dark blue/bright yellow color indicates zero/maximal $A_{\vec{K}}(\varepsilon)$; $S=1$. (a,b) DMI dependence of $A_{\vec{K}}(\varepsilon)$ at $B/J=0.1$ and (c,d) at $B/J=1$. (e,f) Magnetic field dependence of $A_{\vec{K}}(\varepsilon)$ for (e) $D_z/J=0.15$ and (f) $D/J=0.15$. In the chiral magnet (right column), a finite Dirac cone mass is witnessed by the splitting of the quasiparticle peak. 
	}
	\label{fig:spec}
\end{figure}

Next, we focus on the $K$ (and $K'$) point and study the evolution of $A_{\vec{K}} (\varepsilon)$ with $D$ (or $D_z$) and $B$. The results in  Fig.~\ref{fig:spec} corrobarate the above finding. For the achiral magnet (left column in Fig.~\ref{fig:spec}) the quasiparticle peaks stay degenerate but they split for the chiral magnet (right column).

Figures~\ref{fig:spec}(a) and (c) reveal a uniform downwards shift of the degenerate quasiparticle peaks of the achiral magnet at $B/J = 0.1$ and $B/J=1$, respectively. This uniform renormalization is quadratic in DMI, which complies with the observation that the self-energy in Eq.~\eqref{eq:self-energy-at-zeroT} contains products of two interaction vertices, each of which is linear in DMI [cf.~Eqs.~\eqref{eq:V3a}-\eqref{eq:V3d}]. Similarly, the splitting of the Dirac cone in chiral magnets, as depicted in Figs.~\ref{fig:spec}(b) and (d), also grows quadratically with DMI. 

Comparing results at $B/J = 0.1$ and $B/J=1$ (top and middle rows in Fig.~\ref{fig:spec}, respectively), one finds that the quasiparticle line widths are larger in the former case. This is due to the fact that for $B/(JS) \le 1$ the harmonic magnon energies overlap with the two-magnon continuum, rendering spontaneous decays kinematically possible. The momentum-resolved two-magnon density of states (DOS) reads
\begin{align}
	D_{\vec{k}}(\varepsilon) = \frac{1}{N} \sum_{j,j' = \pm} \sum_{\vec{q} \in \text{BZ}} \delta\left( \varepsilon - \varepsilon_{\vec{q},j} - \varepsilon_{\vec{k}-\vec{q},j'} \right)
	\label{eq:twomagnonDOS}
\end{align}  
and is indicated by color in Fig.~\ref{fig:DOS}. One can show that at $\vec{k} = \vec{K}$ (or $\vec{k} = \vec{K}'$) its lower boundary has energy $\varepsilon_{\text{bound}} = 2 J S + 2 B$, where the $2B$ arise from the two-magnon excitations being built from two single-particle excitations, each of which grows with $B$ [cf.~Eq.~\eqref{eq:single-particle-energy}]. Hence, by equating $\varepsilon_\text{D} = \varepsilon_{\text{bound}}$, we find that the harmonic Dirac cone leaves the continuum at the critical field 
\begin{align}
	B_\text{c} = JS.
\end{align}

\begin{figure}
	\centering
	\includegraphics[scale=1]{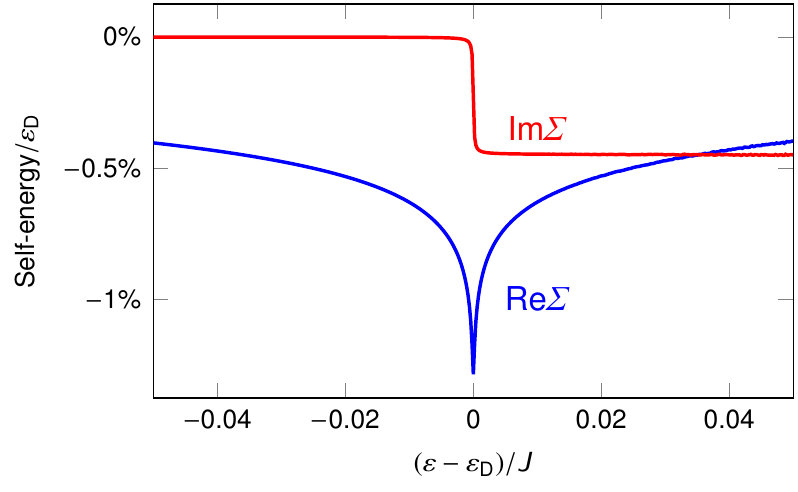}
	\caption{
	Sketch of an element of the self-energy matrix at the lower threshold of the two-magnon continuum for $D/J=0.15$. At $B = B_\text{c} = JS$ and $S=1$, the lower threshold of the two-magnon continuum is exactly at $\varepsilon_\text{D}$ [see Eq.~\eqref{eq:DiracEnergy}], where the imaginary part of the self-energy exhibits a step; $\text{Im} \varSigma$ is zero outside of the continuum ($\varepsilon-\varepsilon_\text{D}<0$) and finite inside ($\varepsilon-\varepsilon_\text{D}>0$). As enforced by the Kramers-Kronig relations, the real part of the self-energy exhibits a logarithmic singularity which appears to be cut only because of a finite numerical linewidth: $0^+ \to 10^{-4} J$. Hence, $\text{Re} \varSigma$ is much larger than $\text{Im} \varSigma$ outside of the continuum. 
	}
	\label{fig:SE}
\end{figure}

The imaginary part of the self-energy follows the two-magnon DOS and, thus, exhibits a step at the lower boundary of the two-magnon continuum, see Fig.~\ref{fig:SE}. According to the Kramers-Kronig relation, a step in the imaginary part translates into a logarithmic singularity of the real part. The explicit expressions for these singularities are derived in Appendix \ref{sec:AppendixLogSingularities}; they read
\begin{subequations}
\begin{align}
	\text{Re}\varSigma^{\alpha \beta}_{\vec{K}}(\varepsilon_{\vec{K}})
    &\sim
    C^{\alpha\beta} J \left( \frac{D}{J} \right)^2 \log \left| \frac{\delta B}{J} \right|,
  %  \dl{ C^{\alpha\beta} J \left( \frac{D}{J} \right)^2 \log \left| \frac{B - B_\text{c}}{J} \right|,}
    \label{eq:real-singularity}
    \\
    \text{Im}\varSigma^{\alpha \beta}_{\vec{K}}(\varepsilon_{\vec{K}})
    &\sim
    C^{\alpha\beta} J \left( \frac{D}{J} \right)^2 \left[ \varTheta \left( \frac{\delta B}{J} \right) -1 \right],
    \label{eq:imag-singularity}
\end{align}
\end{subequations}
where $\delta B = B - B_\text{c}$ and $C^{\alpha\beta}$ is a constant.
(Note that a tiny numerical linewidth of order $10^{-4} J$ was assumed in the numerical calculations, which is why the divergence of the real part is cut in Fig.~\ref{fig:SE}.)
For $B > B_\text{c}$, the imaginary part is zero and the real part still very large due to the singularity. Hence, considerably renormalized magnon quasiparticles with negligible damping are expected for magnetic fields that are slightly detuned from $B_\text{c}$. 

Figures~\ref{fig:spec}(e) and (f) explicitly show this behavior for $S=1$. For fields $B<B_\text{c} = J$, the single-magnon excitations overlap with the two-magnon continuum and the quasiparticle peaks are blurred due to lifetime broadening. However, they become sharp once $B > B_\text{c}$. Since the real part of the self-energy exhibits a logarithmic singularity, the magnon renormalization is particularly prominent at $B \approx B_\text{c}$.
As $B$ increases further, the degree of energy renormalization reduces only logarithmically. The same is true for the splitting observed in Fig.~\ref{fig:spec}(f), which is also due to a real part of a self-energy as we show below.

To further understand the Dirac cone splitting in the chiral magnet, we continue our analytical considerations. The renormalized magnon energies $\tilde{\varepsilon}_{\vec{k},\pm}$ are the solutions of the Dyson equation $ \text{Det} [ \mathcal{G}^{-1}_{\vec{k}}(\varepsilon) ] = 0$, where $\mathcal{G}^{-1}_{\vec{k}}(\varepsilon) = \varepsilon - \mathcal{E}_{\vec{k}} - \varSigma_{\vec{k}}(\varepsilon)$ is the inverse of the interacting matrix Green's function, containing the self-energy matrix
\begin{align}
	\varSigma_{\vec{k}}(\varepsilon) = \begin{pmatrix}
		\varSigma^{--}_{\vec{k}}(\varepsilon) & \varSigma^{-+}_{\vec{k}}(\varepsilon) \\
		\varSigma^{+-}_{\vec{k}}(\varepsilon) & \varSigma^{++}_{\vec{k}}(\varepsilon)
	\end{pmatrix},
\end{align}
as given by Eq.~\eqref{eq:self-energy-at-zeroT}.

At the Dirac points, $\vec{k} = \vec{K}$ or $\vec{k}=\vec{K}'$, the harmonic magnons are in resonance, $\varepsilon_{\vec{K},+} = \varepsilon_{\vec{K},-} = \varepsilon_\text{D}$ and we may invoke the on-shell approximation by replacing the argument of the self-energy by $\varepsilon \to \varepsilon_\text{D}$. Since $\varSigma^{--}_{\vec{K}}(\varepsilon_\text{D}) = \varSigma^{++}_{\vec{K}}(\varepsilon_\text{D}) \equiv \varSigma_\text{diag}(\varepsilon_\text{D})$ and $\varSigma^{-+}_{\vec{K}}(\varepsilon_\text{D})=\varSigma^{+-}_{\vec{K}}(\varepsilon_\text{D}) \equiv \varSigma_\text{off-diag}(\varepsilon_\text{D})$, the real part of the renormalized magnon energies becomes
\begin{align}
	\text{Re}(\tilde{\varepsilon}_{\vec{K},\pm}) \approx \varepsilon_\text{D} + \text{Re}\left[ \varSigma_\text{diag}(\varepsilon_\text{D}) \pm \varSigma_\text{off-diag}(\varepsilon_\text{D}) \right],
	\label{eq:onshell-approx}
\end{align}
from which we read off that the diagonal self-energy $\varSigma_\text{diag}(\varepsilon_\text{D})$ causes a shift and the off-diagonal self-energy $\varSigma_\text{off-diag}(\varepsilon_\text{D})$ a splitting 
\begin{align}
	\Delta \varepsilon = 2 \text{Re} \left[ \varSigma_\text{off-diag}(\varepsilon_\text{D}) \right]
	\label{eq:gap-equation}
\end{align}
of the magnon energy at the Dirac point.

Importantly, the achiral magnet has zero off-diagonal self-energies at the $K$ and $K'$ points, such that there is no splitting, i.e., no Dirac mass. In contrast, the chiral magnet has a nonvanishing $\varSigma_\text{off-diag}$, which explains the splitting observed in Figs.~\ref{fig:tomographic-spectral-function}(b) and \ref{fig:spec}(b), (d), and (f).

\paragraph*{Logarithmic singularity.}
Within the on-shell solution of the Dyson equation in Eq.~\eqref{eq:onshell-approx} the logarithmic singularity encountered in Eq.~\eqref{eq:real-singularity} directly enters the renormalized spectrum. This indicates the failure of lowest-order ($1/S$) perturbative spin-wave expansion, i.e., a violation of the assumption that $|\varSigma_{\vec{k}}(\varepsilon)| \ll \varepsilon_{\vec{k}}$. Although the self-energy scales with $D^2$ and, hence, is hardly larger than $1\%$ of the Dirac energy $\varepsilon_\text{D}$ for $D/J=0.15$ [cf.~Fig.~\ref{fig:SE}], eventually, its real part grows indefinitely as $B \to B_\text{c}$. This defines a small field window about $B_\text{c}$, within which $\mathrm{Re}|\varSigma_{\vec{k}}(\varepsilon)| \gtrsim \varepsilon_\text{D}$ and the lowest-order-in-$1/S$ perturbation theory is inconsistent.
The logarithmic singularity may be cut due to higher-order diagrams which physically account for the two-magnon continuum and, in particular, its thresholds, being renormalized once its building blocks, i.e., the single-particle energies, have been renormalized. Unfortunately, such an analysis is severely hindered by the small-wavelength nature of the problem \cite{Chernyshev2009}.

However, one notes that the spectral function in Figs.~\ref{fig:spec}(e) and (f) does not show divergent quasiparticle peaks at $B = B_\text{c}=J$. This is because it amounts to a non-perturbative solution of the Dyson equation that is no longer on-shell nor strictly of order $1/S$. Alternatively, self-consistency may be enforced by iteratively solving the Dyson equation off-shell to cut the singularity \cite{Chernyshev2009}. We detail the latter idea in Appendix \ref{sec:OffShell}. Importantly, the off-shell solution agrees very well with the spectral function. Since we also find independent numerical evidence for the spectrum by means of exact diagonalization in Sec.~\ref{sec:ExactDiag}, we conclude that lowest-order perturbation theory is sufficient to qualitatively capture the signatures of magnon-magnon interactions.

% ====================================
%  Effective Hamitlonian
% ====================================
\subsection{Interaction-Induced Topology}
\label{sec:ResultsTopology}
We restrict our further analysis to the chiral magnet to obtain insight into the nature of the interaction-induced gap.

\subsubsection{Effective Hamiltonian}
\label{sec:effectiveHam}
In the limit $B \gg B_\text{c}$, the two-magnon continuum ($\sim 2B$) and the single-particle excitations ($\sim B$) are well separated,  rendering damping zero because decays are kinematically forbidden. Moreover, since the distance between the two magnon bands is negligible compared to their distance to the continuum, the difference between $\varSigma_{\vec{k}}(\varepsilon_{\vec{k},+})$ and $\varSigma_{\vec{k}}(\varepsilon_{\vec{k},-})$ is negligible as well. Hence, we set $\varSigma_{\vec{k}}(\varepsilon_{\vec{k},+}) \approx \varSigma_{\vec{k}}(\varepsilon_{\vec{k},-}) \approx \varSigma_{\vec{k}}(\varepsilon_\text{D})$ to construct an effective Hamiltonian
\begin{align}
	\mathcal{H}_{\vec{k}}^\text{eff} 
	&\equiv 
	\mathcal{H}_{\vec{k}} + \mathcal{U}_{\vec{k}} \varSigma^\text{H}_{\vec{k}} (\varepsilon_\text{D}) \mathcal{U}^\dagger_{\vec{k}}
	\label{eq:effective-ham}
\end{align}
where 
$
	\varSigma^\text{H}_{\vec{k}} (\varepsilon_\text{D}) = [\varSigma_{\vec{k}} (\varepsilon_\text{D}) + \varSigma_{\vec{k}}^\dagger (\varepsilon_\text{D})]/2
$ 
is the Hermitian part of the self-energy at $\varepsilon_\text{D}$. Note the appearance of $\mathcal{U}_{\vec{k}}$ in Eq.~\eqref{eq:effective-ham}, necessary for transforming $\varSigma^\text{H}_{\vec{k}} (\varepsilon_\text{D})$, which is evaluated in the eigenbasis, back into the HP basis. $\mathcal{H}_{\vec{k}}^\text{eff}$ describes a two-level system, which may be written as
\begin{align}
	\mathcal{H}_{\vec{k}}^\text{eff} 
	&=
	d^0_{\vec{k}} \sigma_0 + \vec{d}_{\vec{k}} \cdot \vec{\sigma}.
\end{align}
This decomposition in terms of Pauli matrices $\vec{\sigma} = (\sigma_1, \sigma_2, \sigma_3)$ and the unit matrix $\sigma_0$ resembles a spin-$1/2$ Zeeman Hamiltonian with $d^0_{\vec{k}}$ and the components of the ``effective field'' $\vec{d}_{\vec{k}}$ given by 
\begin{subequations}
\begin{align}
	d^0_{\vec{k}} 
	&= 
	3JS + B + \frac{1}{2} \left( \varSigma^{--}_{\vec{k}}+\varSigma^{++}_{\vec{k}}  \right),
	\\
	d_{\vec{k}}^1
	&=
	- \frac{JS}{2} \left( \gamma_{\vec{k}}+\gamma_{-\vec{k}} \right)
	+ \frac{1}{2} \left[ \mathrm{Re}( \tilde{\gamma}_{\vec{k}} ) \left( \varSigma^{--}_{\vec{k}}-\varSigma^{++}_{\vec{k}} \right) \right.
	\nonumber \\
		&\quad \left.
		+\mathrm{i}\mathrm{Im}( \tilde{\gamma}_{\vec{k}} ) \left( \varSigma^{+-}_{\vec{k}}-\varSigma^{-+}_{\vec{k}} \right)
	 	\right],
	\\
	d_{\vec{k}}^2
	&=
	- \mathrm{i} \frac{JS}{2} \left( \gamma_{\vec{k}}-\gamma_{-\vec{k}} \right)
	+ \frac{\mathrm{i}}{2} \left[ \mathrm{Re}( \tilde{\gamma}_{\vec{k}} ) \left( \varSigma^{+-}_{\vec{k}}-\varSigma^{-+}_{\vec{k}} \right) \right.
	\nonumber \\
		&\quad \left.
		+\mathrm{i}\mathrm{Im}( \tilde{\gamma}_{\vec{k}} ) \left( \varSigma^{--}_{\vec{k}}-\varSigma^{++}_{\vec{k}} \right)
	 	\right],
	\\
	d_{\vec{k}}^3
	&= 
	\frac{1}{2} \left( \varSigma^{-+}_{\vec{k}}+\varSigma^{+-}_{\vec{k}} \right),
	\label{eq:massd}
\end{align}
\end{subequations}
where $\tilde{\gamma}_{\vec{k}} = \text{sgn}(\gamma_{\vec{k}})$. For notational ease, we suppressed both the dependence on $\varepsilon_\text{D}$ and the ``H'' label. 

\begin{figure}
	\centering	
	\includegraphics[scale=1]{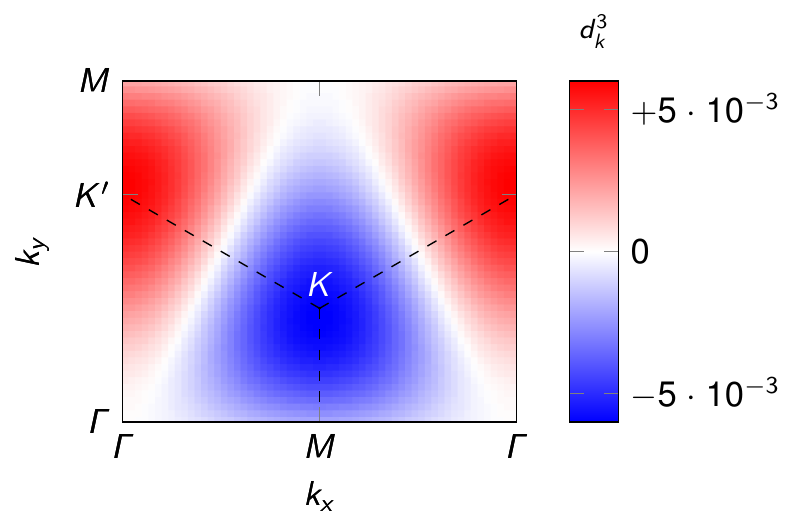}
	\caption{Mass term $d_{\vec{k}}^3$ in the first Brillouin zone at $B/(JS)=4$. Blue/white/red color indicates negative/zero/positive values. Note the opposite sign of $d_{\vec{k}}^3$ at the $K$ and $K'$ points, causing nonzero Chern and winding numbers. Parameters used are $S=1$ and $D/J=0.15$.}
	\label{fig:massterm}
\end{figure}

The eigenvalues of $\mathcal{H}_{\vec{k}}^\text{eff} $ are given by
\begin{align}
	\tilde{\varepsilon}_{\vec{k},\pm}^\text{eff} = d^0_{\vec{k}} \pm | \vec{d}_{\vec{k}}|. \label{eq:energies-effective}
\end{align}
Importantly, at $\vec{k} = \vec{K}$ and $\vec{k} = \vec{K}'$, we obtain $d_{\vec{k}}^1 = d_{\vec{k}}^2 = 0$, but $d_{\vec{k}}^3 \ne 0$, implying a mass gap.\footnote{At the Dirac point $\vec{K}'$, the definition of $d^3_{\vec{k}}$ in Eq.~\eqref{eq:massd} agrees with the definition of $\varSigma_\text{off-diag}$ discussed in Sec.~\ref{sec:ResultsGap}.} Figure \ref{fig:massterm} shows a representative $d^3_{\vec{k}}$ within the entire Brillouin zone, from which it is obvious that the Dirac mass has opposite signs at the $K$ and $K'$ points. 
Hence, the Berry curvature
\begin{align}
	\Omega_{\vec{k},\pm} = \mp \frac{1}{2 d^3_{\vec{k}} }  \vec{d}_{\vec{k}} \cdot \left( \frac{\partial \vec{d}_{\vec{k}}}{\partial k_x} \times \frac{\partial \vec{d}_{\vec{k}}}{\partial k_y} \right)
	\label{eq:Berry-Curvature}
\end{align}
integrates to a nonzero Chern number 
\begin{align}
	C_{\pm} = \frac{1}{2 \pi} \int_\text{BZ} \Omega_{\vec{k},\pm} \text{d}^2 k.
\end{align}
According to the bulk-boundary correspondence \cite{Hatsugai1993, Hatsugai1993a}, the nontrivial winding number \cite{Sticlet2012}
\begin{align}
	w = C_{-} = \frac{1}{2} \left[ \text{sgn}( d_{\vec{K}}^3 ) - \text{sgn}( d_{\vec{K}'}^3 ) \right] = \frac{1}{2} [(-1)-(+1)] = -1
	\label{eq:winding}
\end{align}
suggests a topologically protected chiral edge magnon in finite systems (see Sec.~\ref{sec:ResultsSlab}).

% ====================================
%  Slab calculations
% ====================================
\subsubsection{Interaction-Induced Chiral Edge Magnons}
\label{sec:ResultsSlab}
To verify that the interacting chiral magnet indeed features chiral edge magnons we proceed with simulating a honeycomb-lattice slab with open (periodic) boundary conditions in the $x$ ($y$) direction (cf.~coordinate system in Fig.~\ref{fig:models}). Hence, the edges exhibit a zigzag termination. A width of $12$ honeycomb-lattice unit cells is chosen. Further details on the numerical implementation are given in Appendix \ref{sec:AppendixDetailsSlab}.

\begin{figure}
	\centering
	\includegraphics[scale=1]{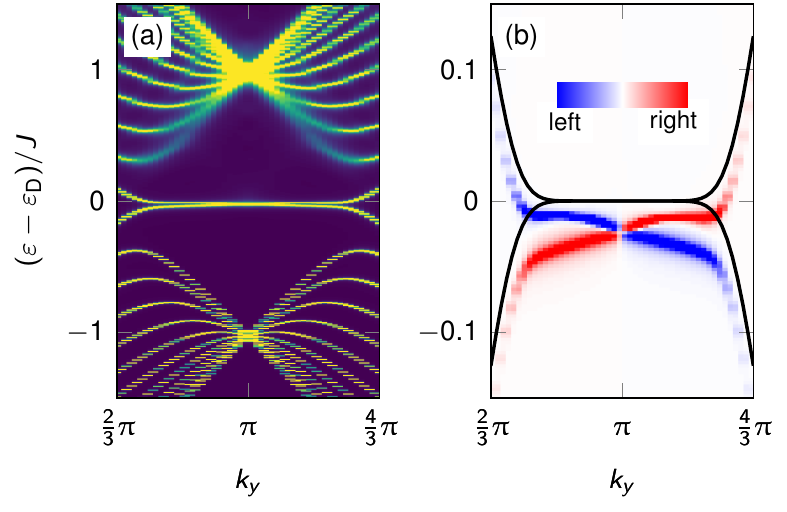}
	\caption{Interaction-corrected magnonic spectral function $A_{k_y}(\varepsilon)$ of a honeycomb-lattice slab of finite width (in $x$ direction) revealing chiral edge states. (a) $A_{k_y}(\varepsilon)$ of a slab with a width of twelve unit cells. Blue/yellow color indicates zero/maximal values; $D/J=0.15$, $S=1$ and $B/(JS)=1$. (b) Position-resolved $A_{k_y}(\varepsilon)$, with blue/red color indicating localization at the left/right edge. Black lines indicate the flat harmonic edge state connecting the edge projections of the Dirac cones for zigzag termination.}
	\label{fig:slab}
\end{figure}

Figure \ref{fig:slab}(a) shows the magnonic spectral function $A_{k_y}(\varepsilon)$ of the interacting slab. $A_{k_y}(\varepsilon)$ almost resembles the harmonic spectrum, exhibiting projections of Dirac cones connected by a flat edge state. For a slab of large width, the Dirac cone projections appear at $k_y = 2\pi/3$ and $k_y = 4\pi/3$. Here, due to finite-size effects, the Dirac cone projections are shifted in reciprocal space towards each other. Nonetheless, the flat edge states are clearly discernible.

Upon zooming into the relevant energy window [see Fig.~\ref{fig:slab}(b)], we find that the edge states do not coincide with the flat harmonic bands (black lines). The edge modes have acquired both a uniform downwards shift and a dispersion due to many-body interactions. A spatially resolved calculation of the spectral function (details in Appendix \ref{sec:AppendixDetailsSlab}) reveals that they are chiral. States on the left (right) edge have positive (negative) slope as dictated by the winding number $w$ in Eq.~\eqref{eq:winding}. This result confirms that interactions open a topologically nontrivial gap, which hosts chiral edge states, and establishes the notion of ``interacting magnon Chern insulators.'' We point out the difference to conventional ``harmonic'' magnon Chern insulators in Sec.~\ref{sec:harmonicVSanharmonic}.

% ====================================
%  Field-induced topological transition
% ====================================

\begin{figure}
	\centering
	\includegraphics[scale=1]{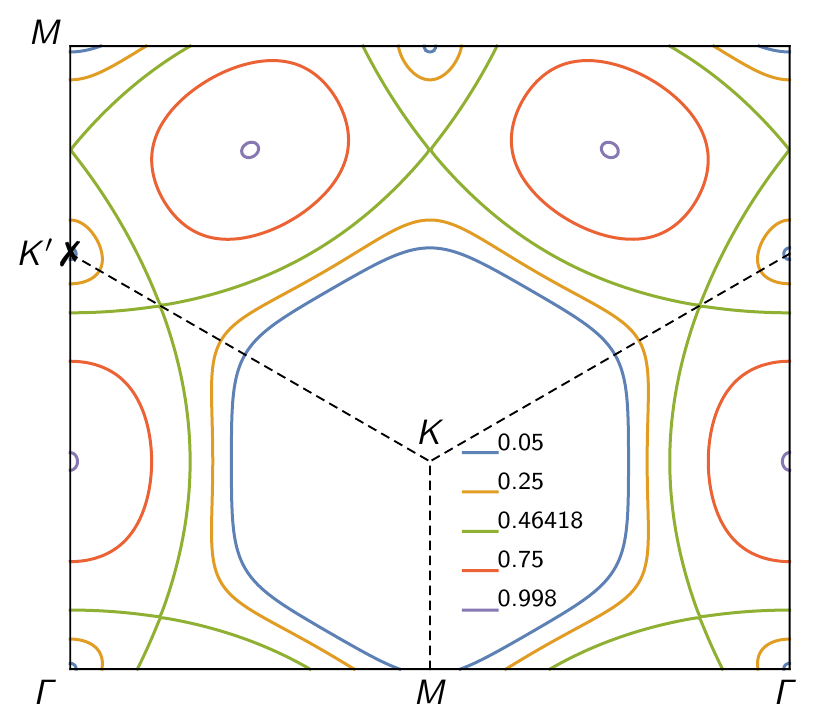}
	\caption{
	Energy-conserving decay contours in reciprocal space for an initial Dirac magnon at the $K'$ point (see \ding{55} symbol) with energy $\varepsilon_\text{D}$. Colored lines correspond to selected values of magnetic field $B/(JS)$ as indicated. At small fields (blue and orange lines) there are three contours, respectively centered around the $\varGamma$, $K'$, and $K$ points. At large fields (red and purple lines), the decay contours are centered about the midpoints of the lines from $\varGamma$ to $K'$. At a critical field, there is a topological transition of the decay contours from the low-field to the high-field case, as indicated by the green lines. This transition corresponds to a saddle point of the two-magnon continuum and, hence, a logarithmic singularity in the imaginary part of the self-energy.
	}
	\label{fig:scattering-lines}
\end{figure}

\subsubsection{Field-Induced Topological Phase Transition}
\label{sec:ResultsTransitionField}
Having collected evidence that $\varSigma_\text{off-diag}(\varepsilon_\text{D}) = \varSigma^{-+}_{\vec{K}}(\varepsilon_\text{D})$---or, analogously, $d^3_{\vec{K}}$ in Eq.~\eqref{eq:massd}---is a useful indicator of topology, we may now explore its behavior in the regime $B < B_\text{c}$, in which the Hermitian effective model \eqref{eq:effective-ham} nominally is a bad approximation both due to lifetime broadening (non-Hermitian contributions) and the self-energy being considerably energy-dependent. Still, crucial information can be extracted, as we explain below. 

First, let us study the energy-conserving decay contours of a Dirac magnon at the $K'$ point, which are depicted in Fig.~\ref{fig:scattering-lines}. At small fields (blue and orange lines), there are two decay channels. The first one is a decay into a magnon close to the $\varGamma$ point and another one close to the initial $K'$ point. The second channel consists of two decay products situated at a ring around the $K$ point. As the field increases, the decay contours grow until they meet for a critical value $B'_\text{c} \approx 0.464 JS$ (green line). Further increase of the field causes a topological transition of the decay contours, which are now centered about the midpoint of the $\varGamma$-to-$K'$ paths. Eventually, the decay contours shrink to a point and disappear at $B = B_\text{c}$ (red and purple lines). Similar contours are obtained for Dirac magnons at the $K$ point; only the role of the $K$ and $K'$ points is interchanged.

\begin{figure}
	\centering
	\includegraphics[scale=1]{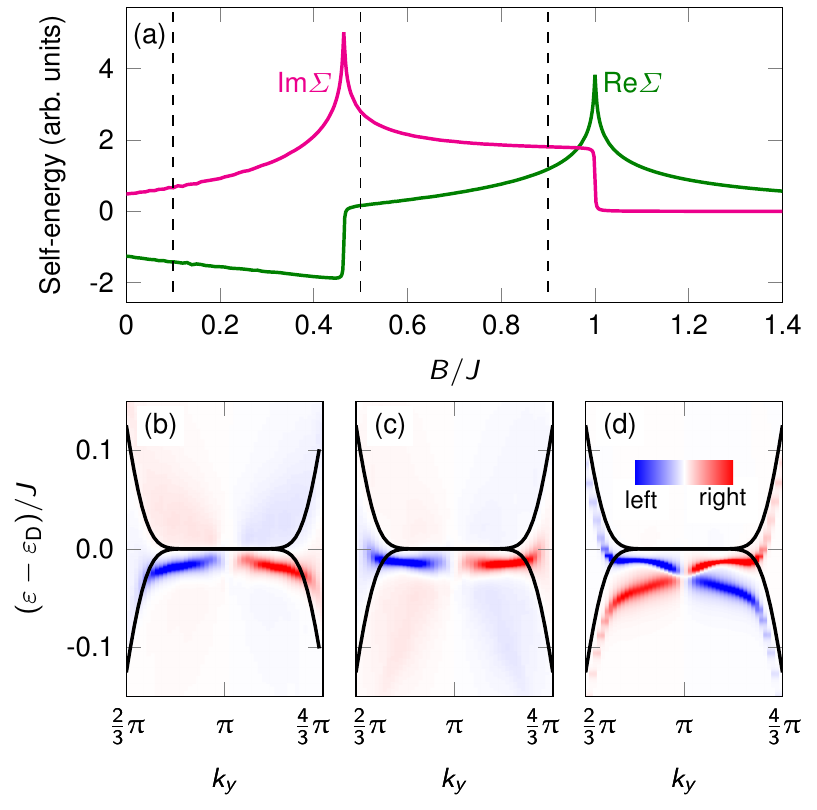}
	\caption{
	Field-induced topological phase transition with damped chiral edge states for $S=1$ and $D/J=0.15$. (a) Real and imaginary part of the self energy $\varSigma^{-+}_{\vec{K}'}(\varepsilon_{\text{D}})$ at the $\vec{K}'$ point and Dirac energy $\varepsilon_{\text{D}}$ in dependence on the magnetic field $B$. At $B=B_\text{c} = J$, the Dirac cone leaves the two-magnon continuum and $\text{Im}\varSigma^{-+}_{\vec{K}'}(\varepsilon_{\text{D}})$ exhibits a step. At $B'_\text{c} \approx 0.464 JS$, there is a saddle point in the two-magnon continuum, associated with the topological transition in the decay contours shown in Fig.~\ref{fig:scattering-lines}. This results in a logarithmic singularity of $\text{Im}\varSigma^{-+}_{\vec{K}'}(\varepsilon_{\text{D}})$ (labeled $\text{Im}\varSigma$ for brevity), which is accompanied by a sign change of $\text{Re}\varSigma^{-+}_{\vec{K}'}(\varepsilon_{\text{D}})$ (labeled $\text{Re}\varSigma$). (b-d) Position resolved $A_{k_y}(\varepsilon)$ of a honeycomb-lattice slab (similar to that in Fig.~\ref{fig:slab}) at field values indicated by dashed lines in panel (a). Blue/red color indicates localization at the left/right edge. Black lines indicate the flat harmonic edge state connecting the edge projections of the Dirac cones for zigzag termination. 
	}
	\label{fig:field-transition}
\end{figure}

A topological transition of the decay contours at $B'_\text{c}$ signals a saddle point of the two-magnon continuum. Hence, the two-magnon DOS exhibits a logarithmic singularity, which is also found in the imaginary part of the self-energy as shown in Fig.~\ref{fig:field-transition}(a) by the purple line. According to the Kramers-Kronig relations, a logarithmic singularity in the imaginary part of the self-energy translates into a sign change of the real part [green line in Fig.~\ref{fig:field-transition}(a)]. For the Hermitian effective model, this finding translates into a sign change of $d^3_{\vec{K}}$ [because $\varSigma_\text{off-diag}(\varepsilon_\text{D})$ flips sign] at $B'_\text{c}$. A global minus sign of the mass term also flips the sign of the winding number $w$ in Eq.~\eqref{eq:winding} and we expect a chirality reversal of the edge states.

Figures \ref{fig:field-transition}(b)--(d) show the spectral function of a honeycomb-lattice slab (similar to the one studied in Sec.~\ref{sec:ResultsSlab}) at three selected values of $B$, which are indicated by dashed lines in Fig.~\ref{fig:field-transition}(a). These field values are chosen (b) below $B'_\text{c}$, (c) almost at $B'_\text{c}$, and (d) above $B'_\text{c}$. The antisymmetric blue-red feature due to the edge states is witnessed to reverse as field increases. The blue (red) state at the left (right) corner has positive (negative) group velocity below $B'_\text{c}$ that turns negative (positive) above $B'_\text{c}$. Right at $B'_\text{c}$ the edge states are flat, complying with the presence of a topological phase transition from positive to negative winding number $w$.

Hence, we demonstrated that the effective model has predictive power even in the limit of overlapping single and two-particle sectors, thereby linking geometric properties of the two-magnon DOS to single-magnon topology. One might conjecture that there is a more general principle at work, rigorously relating the topology of the decay surface to the band topology of single-particle bands, but this is beyond the scope of the present work and is left for future investigation.

% ====================================
%  Magnon Revival
% ====================================
\subsubsection{Revival of magnon topology for strong interactions}
\label{sec:ResultsRevival}
As already mentioned in the introduction, band gaps $\Delta \varepsilon$ and, hence, in-gap chiral states are meaningful terms only if the broadening $\varGamma_{+}$ and $\varGamma_{-}$ of the gapped bands is smaller than the gap. Hence, $\Delta \varepsilon/\sqrt{\varGamma_+ \varGamma_-} > 1$ must hold. In principle, we may consider the following two sources for damping.
 
(1) 
The phenomenological Gilbert damping $\alpha$ leads to a magnon damping $\varGamma_{j,\vec{k}} = \alpha \varepsilon_{j,\vec{k}}$. At the Dirac point, for $S=1$ and $B = B_\text{c}$, we extract $\Delta \varepsilon \approx 2.15 \times J (D/J)^2$ from our numerical data [cf.~Fig.~\ref{fig:spec}(d)]. Hence, we get the approximate relation
\begin{align}
	\frac{\Delta \varepsilon}{\sqrt{\varGamma_+ \varGamma_-}} 
	\approx 
	\frac{\Delta \varepsilon}{\varGamma} 
	\approx
	\frac{1}{\alpha} \left( \frac{D}{J} \right)^2.
\end{align}
and conclude that $(D/J)^2 > \alpha$ must hold. (We set $\varGamma_+ = \varGamma_- = \varGamma = \alpha \varepsilon_\text{D}$.) Since Gilbert damping can be as small as $\alpha < 10^{-3}$ (down to $\alpha \approx 10^{-4}$), already small ratios $D/J$ should cause a well-defined gap.

(2)
As for the magnon-magnon interactions, we have seen in Sec.~\ref{sec:ResultsTransitionField} that for $B<B_\text{c} = JS$ the single-particle states overlap with the continuum, resulting in damping. However, quite counter-intuitively, this reasoning holds only for weak interactions. As shown by Verresen \textit{et al.} \cite{Verresen2019}, who built upon the work of Gaveau and Schulman \cite{Gaveau1995}, strong interactions do \emph{not} fully wipe out the quasiparticle but rather expel it from the continuum. The expelled single-particle state retrieves a long lifetime, which comes at the price of a reduced quasiparticle residue.

In principle, this mechanism can ``revive'' any single-particle state and, in particular, topological magnon band gaps and chiral in-gap states: although being blurred out for weak interactions, the gapped bands (and the chiral edge states) may get expelled from the continuum by strong interactions, thereby suppressing damping.
Let us present this effect for the gapped Dirac magnons of the chiral ferromagnet in Fig.~\ref{fig:models}(b).

\begin{figure}
	\centering
	\includegraphics[scale=1]{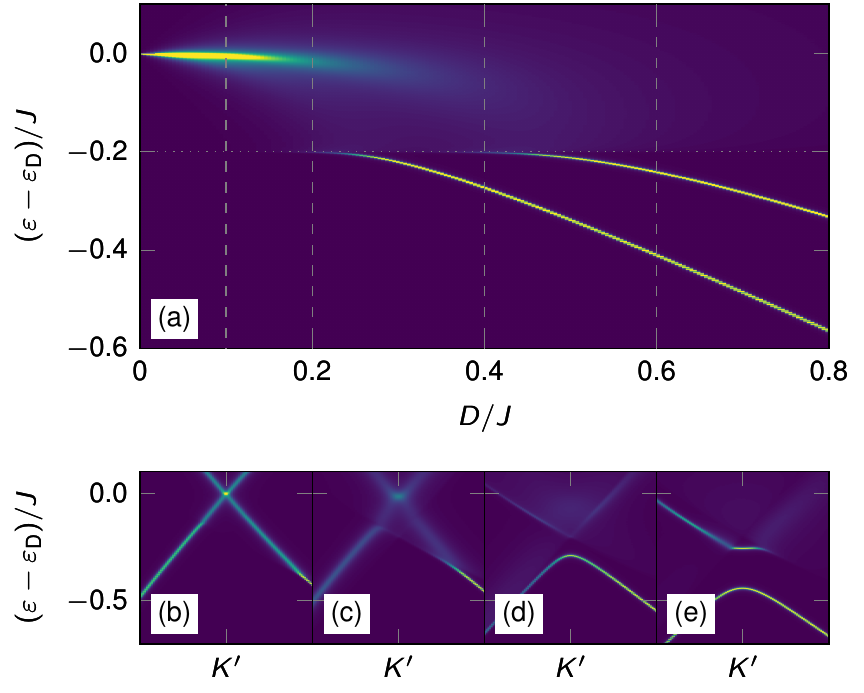}
	\caption{Topological magnon ``revival'' brought about by strong interactions in the chiral magnet for $S=1$ and $B/(JS)=0.8$ (i.e., $B<B_\text{c}$). (a) Spectral function $A_{\vec{K}'}(\varepsilon)$ at the $K'$ point in dependence on DMI $D$. Dark blue/bright yellow color indicates zero/maximal $A_{\vec{K}'}(\varepsilon)$. The dotted horizontal line at $(\varepsilon-\varepsilon_\text{D})/J=-0.2$ indicates the lower threshold of the two-magnon continuum, from which the two quasiparticle peaks are expelled at about $D/J = 0.2$ and $D/J=0.4$, respectively. Vertical dashed lines indicate $D/J$ ratios for which band structures are shown in the lower panels. (b--e) Many-body renormalized magnon band structure along a high-symmetry path in reciprocal space through the $K'$ point for (b) $D/J=0.1$, (c) $D/J=0.2$, (d) $D/J=0.4$, and (e) $D/J=0.6$.}
	\label{fig:revival}
\end{figure} 

Figure \ref{fig:revival}(a) shows how the Dirac cone is expelled from the two-magnon continuum for increasing interactions, i.e., for increasing DMI $D$. With $S=1$, and $B/(JS)=0.8$, the harmonic Dirac cone is energetically located well within the continuum, whose lower threshold is indicated by the horizontal dotted gray line. At $D/J=0$, the Dirac magnon has zero mass and the two quasiparticle peaks are degenerate. Small $D/J$ ratios lead to appreciable lifetime broadening, because decay processes are kinematically allowed. Hence, along a path in reciprocal space through the $K'$ point the spectral function resembles the harmonic Dirac cone with additional broadening [Fig.~\ref{fig:revival}(b)]. For $D/J=0.2$, the damping is already so large that only a very blurred quasiparticle feature can be identified. The Dirac cone shape is almost invisible [Fig.~\ref{fig:revival}(c)]. As $D/J$ increases further, spectral weight is transferred from the continuum to the quasiparticle peak clinging to the lower boundary of the continuum and the lower band of the Dirac cone leaves the continuum [Fig.~\ref{fig:revival}(d)]. Eventually, also the second quasiparticle peak is expelled [Fig.~\ref{fig:revival}(e)]. Being no longer located within the continuum, both peaks are sharp.

% ====================================
%  Thermal topological transition
% ====================================
\subsubsection{Thermal Topological Phase Transition}
\label{sec:ResultsTransitionThermal}
We have already seen in Sec.~\ref{sec:ManyBodyPT} that the most important Feynman diagrams are the bubble diagrams, because the tadpoles integrate to zero and the Hartree term merely compresses the spectrum. Hence, as far as magnon topology at finite temperatures is concerned, we may also restrict to the bubble self-energy in Eq.~\eqref{eq:Sigma-bubble}.
So far, we considered the decay term at zero temperature. At finite temperatures the spontaneous contribution is accompanied by two Bose factors $\rho(\varepsilon_{\vec{q},j},T) + \rho(\varepsilon_{\vec{k}-\vec{q},j'},T)$, which enhance the efficiency of decays. Still, as long as decays are kinematically forbidden ($B>B_\text{c}$), finite temperatures will not cause damping. 

\begin{figure}
	\centering
	\includegraphics[scale=1]{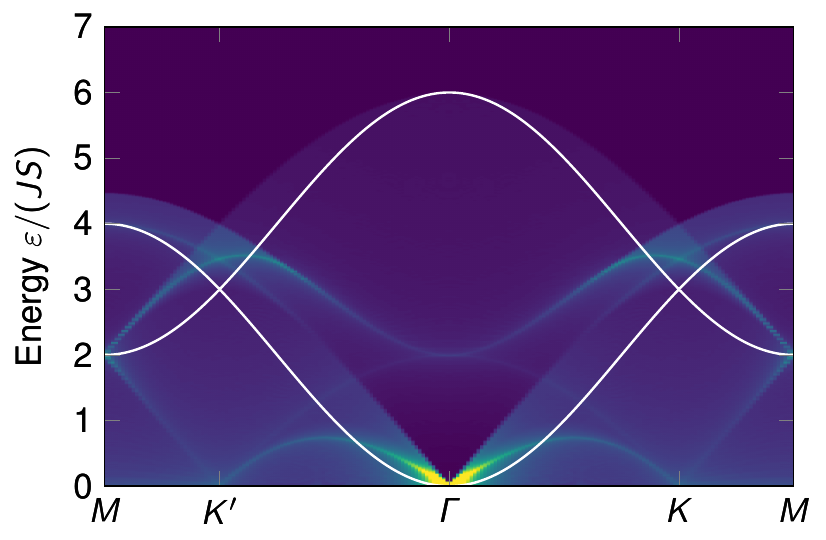}
	\caption{ 
		Collision density of states in honeycomb ferromagnets. At $B=0$, the harmonic single-particle energies [white lines; cf.~Eq.~\eqref{eq:single-particle-energy}] overlap with the collision density of states [color plot; cf.~Eq.~\eqref{eq:collisionDOS}]. Dark blue/yellow color indicates zero/maximal collision density of states.
		For increasing fields, the single-particle energies are shifted upwards in energy by $B$, while the collision continuum is fixed. Hence, for large enough fields, the single-particle bands are shifted past the continuum. In particular, the Dirac cones at the $K$ and $K'$ points leave the continuum at $B = B_\text{c} = JS$, similar to the two-magnon density of states in Fig.~\ref{fig:DOS}.
	}
	\label{fig:DOScollisions}
\end{figure}

A qualitatively new contribution comes from the thermally activated collision term which is proportional to a difference of Bose functions: $\rho(\varepsilon_{\vec{q},j},T) - \rho(\varepsilon_{\vec{k}+\vec{q},j'},T)$. It may contribute to damping, if the single-particle energies overlap with the collision DOS
\begin{align}
	C_{\vec{k}}(\varepsilon) = \frac{1}{N} \sum_{j,j' = \pm} \sum_{\vec{q} \in \text{BZ}} \delta \left( \varepsilon + \varepsilon_{\vec{q},j} - \varepsilon_{\vec{k}+\vec{q},j'} \right),
	\label{eq:collisionDOS}
\end{align}  
which is depicted in Fig.~\ref{fig:DOScollisions}.
However, one verifies that the single-particle energies leave the collision DOS at the same critical field $B_\text{c}$ at which they also leave the two-magnon DOS. Hence, for $B>B_\text{c}$, damping is strongly suppressed at all temperatures (within $1/S$ perturbation theory). The most relevant influence of temperature is then found in its contribution to the real part of the self-energy, with logarithmic singularities arising both from the (lower) two-magnon as well as the (upper) collision DOS thresholds.

\begin{figure}
	\centering
	\includegraphics[scale=1]{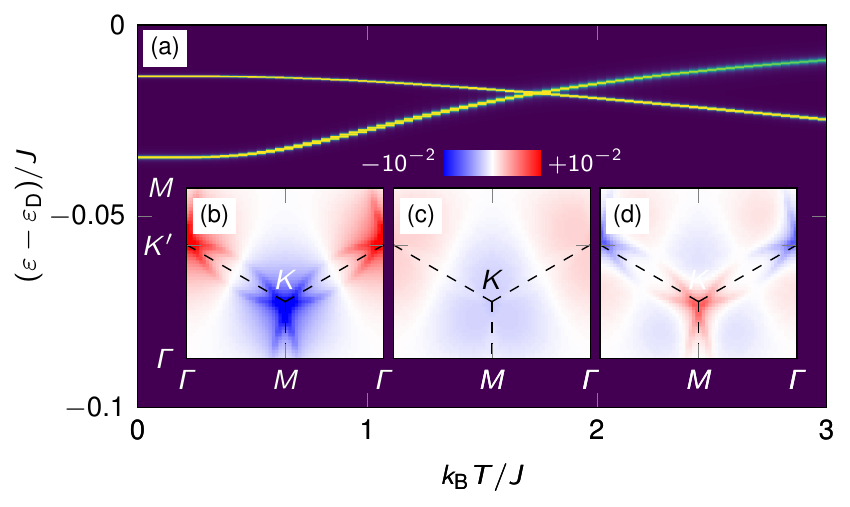}
	\caption{ 
		Temperature dependence of the interaction-induced magnonic Dirac mass at $B/(JS)=1.1$ for $S=1$ and $D/J=0.15$. (a) Magnonic spectral function $A_{\vec{K}}(\varepsilon)$ in dependence on temperature. Dark blue/bright yellow color indicates zero/maximal $A_{\vec{K}}(\varepsilon)$. (b,c,d) Mass term $d^3_{\vec{k}}$ in the hexagonal unit cell at $k_\text{B} T/J=0$, $k_\text{B} T/J=1.8$, and $k_\text{B} T/J=3$, respectively.
	}
	\label{fig:temp}
\end{figure}

Figure~\ref{fig:temp}(a) shows the temperature dependence of the magnonic spectral function $A_{\vec{K}}(\varepsilon)$ for $B=1.1 JS >B_\text{c}$ and $S=1$ at the Dirac cones, taking only bubble diagrams into account. The two quasiparticle peaks stay well defined, in agreement with the aforementioned kinematic reasoning. Just below $k_\text{B} T/ J \approx 2$, there is a band gap closing and reopening. Within the effective model \eqref{eq:effective-ham}, this transition is associated with a vanishing of the off-diagonal self-energies and, hence, of the mass term $d_{\vec{k}}^3$ in Eq.~\eqref{eq:massd}. One verifies that $d_{\vec{k}}^3$ flips sign at the $K$ and $K'$ points, as shown in Figs.~\ref{fig:temp}(b)-(d), leading to a temperature-driven topological phase transition from negative chirality ($w=-1$) to positive chirality ($w=+1$).

To get some intuition, why such a transition can happen, let us ignore the interaction vertices in the self-energy [Eq.~\eqref{eq:Sigma-bubble}]. Then, decays and collisions are fully characterized by the respective DOS. Note that the two-magnon DOS in Eq.~\eqref{eq:twomagnonDOS}, the single-particle energies in Eq.~\eqref{eq:single-particle-energy}, and the collision DOS in Eq.~\eqref{eq:collisionDOS} respectively grow with $2B$, $1B$, and $0B$. At the $K$ and $K'$ points, for $B=B_\text{c}=JS$, the lower threshold of the two-magnon DOS, the Dirac cone in Eq.~\eqref{eq:DiracEnergy}, and the upper threshold of the collision DOS coincide in energy. Thus, for $B>B_\text{c}$, the two-magnon (collision) DOS is energetically above (below) the Dirac cone, whence it follows that
\begin{subequations}
\begin{align}
	\varepsilon_\text{D} - \varepsilon_{\vec{q},j} - \varepsilon_{\vec{K}-\vec{q},j'} &< 0, \\
	\varepsilon_\text{D} + \varepsilon_{\vec{q},j} - \varepsilon_{\vec{K}+\vec{q},j'} &> 0,
\end{align}
\end{subequations}
(for all $\vec{q}$) and the two denominators in Eq.~\eqref{eq:Sigma-bubble} have opposite sign. Consequently, decays and collisions cause mass terms of opposite sign. At low temperatures, when collisions are frozen out, decays dominate. However, thermally activated collisions may take eventually over, causing the topological transition. 

A comment on the consistency of spin-wave theoy at large temperatures is due. Above, we pushed the lowest-order anharmonic spin-wave theory to temperatures $k_\text{B} T > J$. In this limit, temperature-enabled higher-order $1/S$ corrections can be expected to alter our findings. For example, at order $1/S^2$, sunset diagrams, as obtained within second-order many-body perturbation theory in $\hat{H}_4$, are the leading source of damping \cite{Pershoguba2018}. The resulting magnon lifetimes will blur the quasiparticle peaks in Fig.~\ref{fig:temp}. Whether or not the topological transition in the spectrum will be visible depends again on the ratio of the spectral gap to the damping (cf.~discussion in Sec.~\ref{sec:ResultsRevival}).
Moreover, thermal fluctuations eventually destroy the magnetic order. In Sec.~\ref{sec:ResultsTransport}, we estimate the region in parameter space spanned by $B$ and $T$ within which field freezing efficiently counteracts thermal fluctuations. 

Thus, the main insight of our analysis of temperature effects is that thermal fluctuations habor, in principle, the potential to cause topological phase transitions. For more quantitative predictions, higher anharmonic spin-wave interactions have to be taken into account.

% ====================================
%  Full Phase Diagram / Transport
% ====================================
\subsubsection{Interaction-Induced Topological Phase Diagram and Transverse Transport}
\label{sec:ResultsTransport}
\begin{figure}
	\centering
	\includegraphics[scale=1]{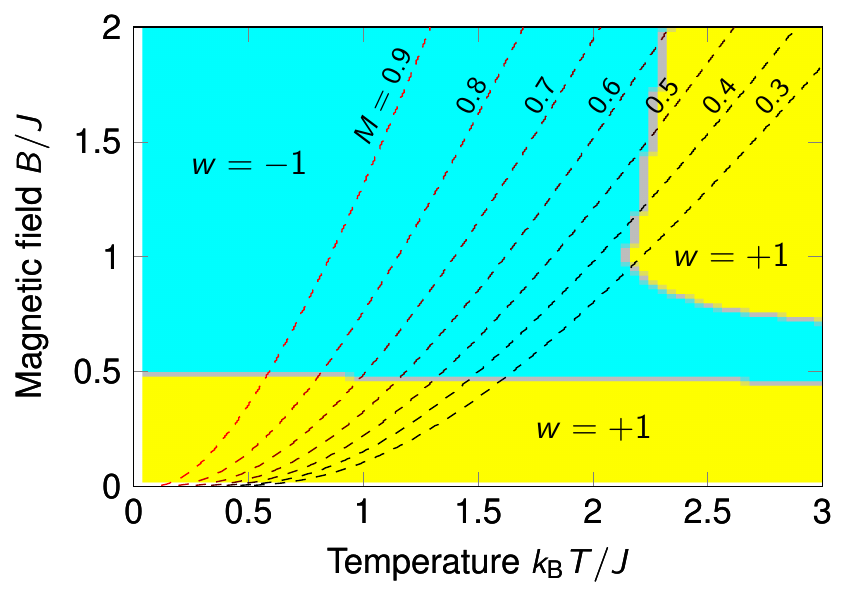}
	\caption{ 
		Interaction-induced magnonic topological phase diagram of the chiral honeycomb ferromagnet in Fig.~\ref{fig:models}(b) as extracted from the effective model Hamiltonian \eqref{eq:effective-ham}; $S=1$ and $D/J=0.15$. Both a change in magnetic field $B$ as well as in temperature $T$ causes topological phase transitions. Dashed lines indicate the relative magnetization $M$ [see Eq.~\eqref{eq:magnetization}; note $M(T=0)=1$], as evaluated within linear spin-wave theory.
	}
	\label{fig:diagram}
\end{figure}

Armed with the knowledge of both field-induced (cf.~Sec.~\ref{sec:ResultsTransitionField}) and temperature-induced topological phase transitions (cf.~Sec.~\ref{sec:ResultsTransitionThermal}), we can now map out the full topological phase diagram shown in Fig.~\ref{fig:diagram}. To that end, we assume the on-shell approximation and extract the winding number $w$ from the sign of the gap in Eq.~\eqref{eq:gap-equation}. We point out that this is an approximation, the accuracy of which may be judged by comparing the gap closing in Fig.~\ref{fig:temp} (below $k_\text{B} T / J =2$; obtained from the spectral function) with the temperature-induced topological phase transition in Fig.~\ref{fig:diagram} (above $k_\text{B} T / J =2$; obtained within the effective model). Despite this slight quantitative disagreement all qualitative features are captured. We find that there are no topologically trivial phases ($w=0$).

Since finite temperatures try to destroy magnetic order, we have to make sure that the relative magnetization (per spin)
\begin{align}
	M(T) \approx 1 - \frac{1}{N} \sum_{\vec{k}} \left[ \rho(\varepsilon_{\vec{k},-},T) + \rho(\varepsilon_{\vec{k},+},T) \right],
	\label{eq:magnetization}
\end{align}
evaluated within linear spin-wave theory, is well above zero. Dashed lines in Fig.~\ref{fig:diagram} indicate lines of constant $M$. 
We find that the temperature-induced topological transition happens at temperatures, at which $M$ is already considerably reduced. Hence, large fields $B/(k_\text{B} T) > 1$ are necessary to freeze the magnetization and to appreciate this transition. 

We note that such a topological phase diagram can never be obtained within the noninteraction theory of ferromagnetic topological magnon insulators. This is because (i) the influence of temperature on the spectrum escapes the harmonic theory as it may only be incorporated as an effective scaling and (ii) external magnetic fields cause but a mere uniform energetic shift to the single-particle excitations [cf.~Eq.~\eqref{eq:single-particle-energy}] without influencing topology. In the language of the effective Hamiltonian, fields only enter the irrelevant constant shift $d_{\vec{k}}^0$ [cf.~Eq.~\eqref{eq:energies-effective}]. Hence, Fig.~\ref{fig:diagram} summarizes the possibilities of interaction-induced topological phase transitions.

To search for experimental signatures of these transitions we propose to perform transport measurements.
We recall that time-reversal breaking is one of the two necessary ingredients for transverse particle transport, the other being compatibility of the magnetic point group with ferromagnetism. With both requirements met by the chiral magnet, we expect anomalous transverse magnon transport at finite temperatures, namely, Hall, Nernst, and Righi-Leduc effects (also called thermal Hall effect). 
They are respectively quantified by off-diagonal elements of the magnetization conductivity $\sigma$, magnetothermal conductivity $\Upsilon$, and thermal conductivity $\kappa$. These conductivities relate magnetization currents $\vec{j}$ and heat currents $\vec{q}$ to gradients in magnetic fields, $\vec{\nabla} B$, and temperature, $\vec{\nabla} T$. The constitutive equations read
\begin{subequations}
\begin{align}
	\vec{j} &= L^{(0)} \vec{\nabla} B - L^{(1)} T^{-1} \vec{\nabla} T ,
	\\
	\vec{q} &= L^{(1)} \vec{\nabla} B - L^{(2)} T^{-1} \vec{\nabla} T ,
\end{align}
\end{subequations}
and the conductivities are defined as $\sigma = L^{(0)}$, $\Upsilon = L^{(1)}/T$, and
\begin{align}
	\kappa = \frac{1}{T} \left( L^{(2)} - L^{(1)} (L^{(0)})^{-1} L^{(1)} \right). \label{eq:thermal-cond}
\end{align}
Notice that $\kappa$ is composed from two contributions, with the second one deriving from a particle backflow, establishing a magnonic Wiedemann-Franz law\footnote{The magnonic Lorenz number reads $\mathcal{L} \equiv \kappa_{xx}/(T \sigma_{xx}) = C k^2_\text{B}/(g \mu_\text{B})^2$, with g-factor $g$, Bohr's magneton $\mu_\text{B}$ and a numerical constant $C$, whose value depends on the dimension \cite{Nakata2015,Nakata2016}. By restricting the Boltzmann transport theory laid out in Refs.~\onlinecite{Nakata2017QSHE} and \onlinecite{Mook2018} to two dimensions, we obtain $C=2$.} both for longitudinal \cite{Nakata2015} as well as transverse transport \cite{Nakata2016}.
The transport tensors $L^{(i)}$ with $i=0,1,2$ are $2\times 2$ matrices, whose elements read \cite{Katsura2010, Matsumoto2011, Matsumoto2011a, Lee2015, Han2016ax, Cheng2016, Zyuzin2016, Nakata2016, Nakata2017QSHE, wang2018anomalous, Mook2018, Kim2019Ferri}
\begin{subequations}
\begin{align}
	L^{(i)}_{xy}(T) &= -\frac{(g \mu_\text{B})^{2-i}}{\hbar} (k_\text{B} T )^i  \sum_{j=\pm} \int_\text{BZ} c_i[ \rho( \varepsilon_{\vec{k},j}, T ) ] \Omega_{\vec{k},j} \, \mathrm{d}^2 k, \label{eq:transportL}
	\\
	L^{(i)}_{\mu\mu}(T) &= (g \mu_\text{B})^{2-i} \sum_{j=\pm} \int_\text{BZ} \tau_{\vec{k},j} \left( v^\mu_{\vec{k},j} \right)^2 \varepsilon_{\vec{k},j}^i \left( -\frac{\partial \rho(\varepsilon_{\vec{k},j},T)}{\partial \varepsilon} \right)  \, \mathrm{d}^2 k, \label{eq:transportL-long}
\end{align}
\end{subequations}
where $\mu = x,y$; moreover, note that $L^{(i)}_{yx}(T)=-L^{(i)}_{xy}(T)$. We introduced the g-factor $g$, Bohr magneton $\mu_\text{B}$, the transport relaxation time $\tau_{\vec{k},j}$, the group velocity $v^\mu_{\vec{k},j}$, and the weights
\begin{subequations}
\begin{align}
	c_0(x) &= x, \\
	c_1(x) &= (1+x) \text{ln} (1+x) - x \text{ln} (x), \\
	c_2(x) &= (1+x) \text{ln}^2 [(1+x)/x] - \text{ln}^2 (x) - 2 \text{Li}_2 (-x).
\end{align} 
\end{subequations}

First, we study the off-diagonal elements $L^{(i)}_{xy}(T)$ in Eq.~\eqref{eq:transportL} and note that we restrict our analysis to intrinsic contributions to transport. Thus, the transverse conductivities are related to the magnon Berry curvature $\Omega_{\vec{k},\pm}$ in Eq.~\eqref{eq:Berry-Curvature} of the effective Hamiltonian $\mathcal{H}_{\vec{k}}^\text{eff}$. Note that temperature enters both explicitly via the thermal weights $c_i(x)$ as well as implicitly because $\mathcal{H}_{\vec{k}}^\text{eff}$ and $\Omega_{\vec{k},\pm}$ inherit the temperature dependence of the self-energy in Eq.~\eqref{eq:Sigma-bubble}.

\begin{figure}
	\centering
	\includegraphics[scale=1]{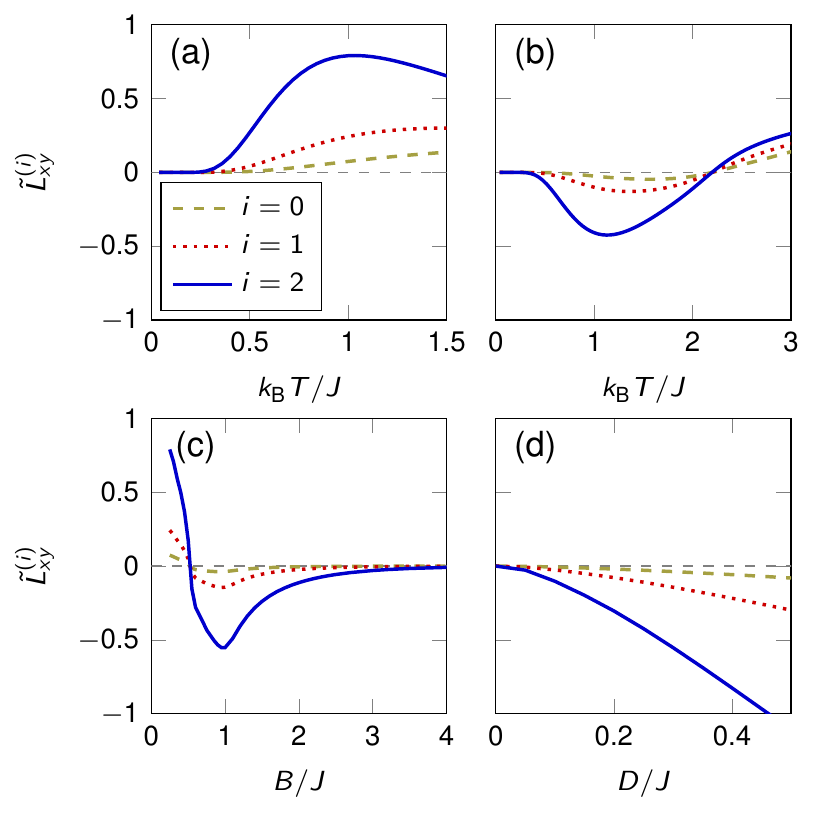}
	\caption{ 
		Interaction-induced transverse conductivities $\tilde{L}_{xy}^{(i)}$ [see Eq.~\eqref{eq:Ltilde}] for $S=1$. (a) Low-field temperature dependence of $\tilde{L}_{xy}^{(i)}$; parameters read $B/J = 0.25$ and $D/J=0.3$. (b) High-field temperature dependence of $\tilde{L}_{xy}^{(i)}$; parameters read $B/J = 1.2$ and $D/J=0.3$. (c) Field-dependence of $\tilde{L}_{xy}^{(i)}$ at $k_\text{B} T/J = 1$, with $D/J=0.3$. (d) Dependence of $\tilde{L}_{xy}^{(i)}$ on DMI $D$ at $k_\text{B} T/J = 1$ and $B/J=1$. Notice the zero crossings that are associated with topological phase transitions in the magnon spectrum: the zero crossing in (b)/(c) is associated with the thermal phase transition/field-induced transition studied in Sec.~\ref{sec:ResultsTransitionThermal}/Sec.~\ref{sec:ResultsTransitionField}. 
	}
	\label{fig:conductivities}
\end{figure}

Figure \ref{fig:conductivities} shows the dimensionless off-diagonal entries of the transport tensors,
\begin{align}
	\tilde{L}^{(i)}_{xy} = -\frac{\hbar L^{(i)}_{xy}}{ (g \mu_\text{B})^{2-i} (k_\text{B} T )^i}
	=
	\sum_{j=\pm} \int_\text{BZ} c_i[ \rho( \varepsilon_{\vec{k},j}, T ) ] \Omega_{\vec{k},j} \, \mathrm{d}^2 k,
	\label{eq:Ltilde}
\end{align}
in dependence on selected parameters. At low fields, $\tilde{L}^{(i)}_{xy}$ is positive over a large temperature window [cf.~Fig.~\ref{fig:conductivities}(a)], which complies with the positive-winding low-field phase in Fig.~\ref{fig:diagram} (lower yellow phase with $w=+1$). In contrast, at higher fields, $\tilde{L}^{(i)}_{xy}$ is negative but the temperature-induced transition (from the cyan $w=-1$ to the yellow $w=+1$ phase in Fig.~\ref{fig:diagram}) eventually flips the sign of the dimensionless conductivities [cf.~Fig.~\ref{fig:conductivities}(b)]. Another sign change is found for a magnetic field sweep, as shown in Fig.~\ref{fig:conductivities}(c). Large fields eventually freeze out magnonic transport because the single-particle bands are shifted towards high energies, suppressing thermal occupation. Finally, an increase in DMI $D$ also causes an increase in transverse conductivities [cf.~Fig.~\ref{fig:conductivities}(d)].

The sign changes of $L^{(0)}_{xy}$ and $L^{(1)}_{xy}$ may be detected by transverse spin transport experiments \cite{Shiomi2017}. However, the most frequently measured signature of transverse magnon transport are heat currents \cite{Onose2010, Ideue12, Hirschberger2015}. According to Eq.~\eqref{eq:thermal-cond}, $L^{(2)}_{xy}$ is not sufficient for a quantitative prediction because the off-diagonal element of $L^{(1)} (L^{(0)})^{-1} L^{(1)}$ has to be subtracted. In contrast to electrons, whose internal energy scale given by the Fermi energy renders this correction negligible at low temperatures, bosons do not come with such a scale \cite{Nakata2015, Nakata2016}. Hence, we need the full $L^{(0)}$ and $L^{(1)}$ tensors, in particular, their diagonal elements given by Eq.~\eqref{eq:transportL-long}. We approximate the transport relaxation time by the magnon lifetime due to Gilbert damping $\alpha$, i.e., $\tau_{\vec{k,j}} \approx \hbar/(2\alpha \varepsilon_{\vec{k,j}})$. Setting $\alpha = 10^{-3}$ results in $\kappa_{xy} / \kappa_{xx} \sim 10^{-3}$, a ratio typically found in experiments \cite{Onose2010, Hirschberger2015}. Moreover, to obtain more wieldy units, we study the three-dimensional thermal conductivity $\kappa^\text{3D} = \kappa / \ell$, which is given as the ratio of conductivity per layer, $\kappa$, and a typical interlayer spacing $\ell = \unit[1]{nm}$. Thus, the units of the thermal conductivity are $[\kappa^\text{3D}]=\unit{\frac{W}{Km}}$.

\begin{figure}
	\centering
	\includegraphics[scale=1]{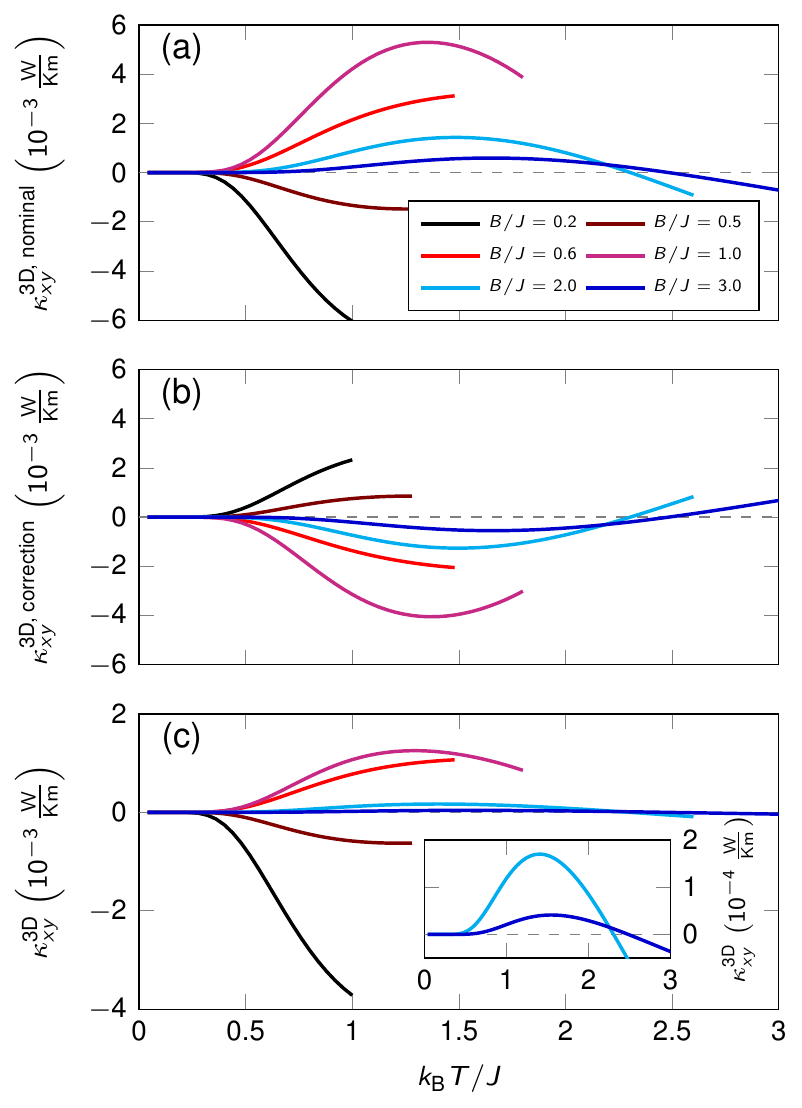}
	\caption{ 
		Interaction-induced transverse thermal conductivity $\kappa_{xy}^\text{3D} = \kappa_{xy}^\text{3D, nominal} + \kappa_{xy}^\text{3D, correction}$ [see Eq.~\eqref{eq:kappa3d}] in dependence on temperature for selected magnetic field values $B/J$. 
		(a) The nominal contribution $\kappa_{xy}^\text{3D, nominal}$ shows the same features as $\tilde{L}^{(2)}_{xy}$ [cf.~Fig.~\ref{fig:conductivities}], i.e., a sign change with both magnetic field and temperature.
		(b) The correction term $\kappa_{xy}^\text{3D, correction}$ has a sign opposite to that of the nominal contribution and is of the same magnitude. 
		(c) The total thermal Hall conductivitiy $\kappa_{xy}^\text{3D}$ exhibits features identical to that of $\kappa_{xy}^\text{3D, nominal}$ but its magnitude is reduced. The inset shows $\kappa_{xy}^\text{3D}$ for fields $B/J > 1$.
		For each magnetic field value, we define an upper threshold temperature $T^\star$ by $M(T^\star) = 0.5$ [see Eq.~\eqref{eq:magnetization} and Fig.~\ref{fig:diagram}]. For $T<T^\star$, the spin-wave theory carried out in this manuscript is expected to be a good approximation. In contrast, temperatures $T>T^\star$ require higher-order spin-wave corrections. 
		Parameters are $S=1$ and $D/J = 0.15$.
	}
	\label{fig:kappa}
\end{figure}

We decompose the transverse thermal conductivity as 
\begin{align}
	\kappa_{xy}^\text{3D} = 
	\underbrace{\frac{L^{(2)}_{xy}}{T \ell}}_{\kappa_{xy}^\text{3D, nominal}} +
	\underbrace{\frac{1}{T \ell}\left(-L^{(1)} (L^{(0)})^{-1} L^{(1)}\right)_{xy}}_{\kappa_{xy}^\text{3D, correction}}.
	\label{eq:kappa3d}
\end{align}
Figures \ref{fig:kappa}(a,b) show the two contributions to the thermal conductivity and Fig.~\ref{fig:kappa}(c) their sum for a realistic ratio of $D/J=0.15$. Importantly, the sign of the correction term [Fig.~\ref{fig:kappa}(b)] is opposite to that of the nominal term [Fig.~\ref{fig:kappa}(a)], which is a result of the backflow current. Due to the missing internal energy scale of magnons (zero chemical potential) both contributions have a similar magnitude. However, since the backflow cannot be larger than what drives the magnonic density imbalance in the first place, $|\kappa_{xy}^\text{3D, nominal}|> |\kappa_{xy}^\text{3D, correction}|$ holds, and the total thermal Hall conductivity $\kappa_{xy}^\text{3D}$ [Fig.~\ref{fig:kappa}(c)] has the same sign as the nominal contribution. Consequently, a measurement of $\kappa_{xy}^\text{3D}$ probes the unconventional topological phase transitions associated with magnon-magnon interactions. Sign changes of $\kappa_{xy}^\text{3D}$ are expected both for the field-induced as well as the temperature-induced transition.
We find that $\kappa_{xy}^\text{3D}$ is of the order of $\unit[10^{-4}]{\frac{W}{Km}}$ to $\unit[10^{-3}]{\frac{W}{Km}}$. Hence, the interaction-induced thermal Hall conductivity can be as large as that obtained within the free theory \cite{Mook14a,McClarty2018} and, in particular, as that frequently observed in experiments \cite{Onose2010, Ideue12, Hirschberger2015}. It should be considered when interpreting transport experiments.

We reiterate that we concentrated on intrinsic transport and worked with the effective Hamiltonian constructed in Sec.~\ref{sec:effectiveHam}. If skew scattering and side jump effects due to disorder can be neglected, this approximation is expected to capture the relevant physics. Still, future theoretical studies may develop a full many-body theory of transverse transport effects. In principle, one would not only expect intrinsic contributions due to the magnon Berry curvature but also skew-scattering-like extrinsic contributions due to asymmetric three-magnon scattering \cite{Verba2019}.

% ====================================
%  Exact diagonalization
% ====================================
\section{Ultimate Quantum Limit: Exact diagonalization for spin $1/2$}
\label{sec:ExactDiag}
So far, we relied on a large-$S$ approximation, which we pushed to $S=1$, a procedure which may be met with scepticism. To confirm the validity of this procedure, we push our analysis even further down to the ultimate quantum limit of $S=1/2$ and compare the results to those obtained from exact diagonalization (ED) of the chiral magnet.
Simulations are performed on a cluster of $12$ honeycomb unit cells ($24$ spins in total) with tilted periodic boundary conditions.

\begin{figure}
	\includegraphics[scale=0.9]{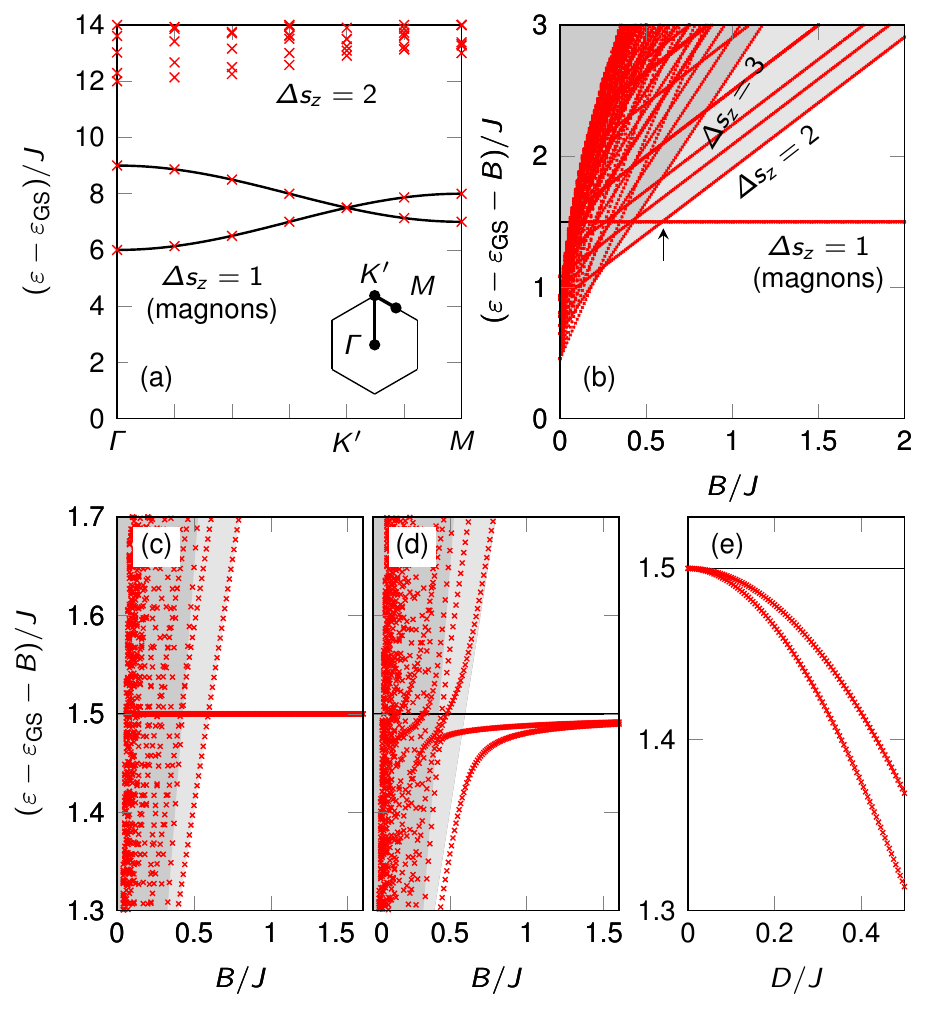}
	\caption{Spectrum of the chiral magnet obtained from exact diagonalization (ED) in the quantum limit of $S=1/2$. Energies are given with respect to the ground state energy $\varepsilon_\text{GS}$. (a) Momentum dependence of the spectrum for $D/J=0$ at $B/J = 6$, causing an energetic separation between $\Delta s_z = 1$ excitations (magnons) and higher-spin excitations. Black lines indicate the spin-wave spectrum given in Eq.~\eqref{eq:single-particle-energy} and the inset shows the hexagonal Brillouin zone with a highlighted path connecting high-symmetry points. (b) Field dependence of the spectrum at the $K$ (or $K'$) point, at which the $\Delta s_z = 1$ excitations (magnons) are degenerate for $D/J=0$. Note that the ordinate shows $(\varepsilon-\varepsilon_\text{GS} -B)/J$, such that $\Delta s_z = 1$ excitations appear as a horizontal line and only higher-spin excitations (indicated by gray continua) shift with increasing field. For fields just above $B_\text{c}=JS=0.5$ (see arrow), the $\Delta s_z = 1$ excitations cut into the $\Delta s_z = 2$ excitations. Eigenstates with different spin quantum numbers do not hybridize due to the absence of DMI. (c) Zoom into the energy and field window, within which the $\Delta s_z = 1$ and $\Delta s_z = 2$ excitations overlap. (d) Same as (c) but with finite DMI $D/J=0.15$, causing spin nonconservation and a hybridization between different spin sectors. Importantly, the formerly degenerate $\Delta s_z = 1$ excitations get gapped. This gap increases the closer they get to the $\Delta s_z = 2$ excitations. (e) Splitting of the $\Delta s_z = 1$ excitations in dependence on DMI at $B/J=1$. For all simulations, a cluster of $24$ spins was considered. In (b,c,d), the lowest $100$ eigenvalues are plotted.} 
	\label{fig:ED}
\end{figure}

For $D=0$, the $z$-component of spin is conserved and the eigenstates can be labeled by their spin $\Delta s_z = 1, 2, 3, \ldots$, with $\Delta s_z = 1$ being magnons, $\Delta s_z = 2$ being two-magnon states, and so on. Large fields energetically separate distinct $\Delta s_z$ sectors, facilitating a comparison with spin-wave theory. Figure \ref{fig:ED}(a) shows the eigenspectrum (red crosses) as obtained from ED at $B/J=6$. The states lowest in energy are $\Delta s_z = 1$ states that excellently agree with the single-particle magnon dispersion obtained within spin-wave theory [Eq.~\eqref{eq:single-particle-energy} for $S=1/2$]. In particular, the two $\Delta s_z=1$ states are degenerate at the $K$ and $K'$ points.
At these points, one finds that the $\Delta s_z \ge 2$ states cross the single-magnon states for sufficiently low fields, as depicted in Fig.~\ref{fig:ED}(b). At fields slightly above $B_\text{c}=JS =0.5$, the single-magnon states cut into the two-magnon manifold. However, due to spin conservation there is no hybridization between different spin sectors, as can be verified in Fig.~\ref{fig:ED}(c).

Hybridization between spin sectors is brought about by finite DMI ($D/J=0.15$) that violates spin conservation. As depicted in Fig.~\ref{fig:ED}(d), the Dirac magnons split and get downwards renormalized as they are approached by two-magnon states. The splitting increases quadratically with growing $D/J$, as shown in Fig.~\ref{fig:ED}(e) for $B/J=1$.

Overall, we find very good qualitative agreement with the anharmonic spin-wave theory, e.g., with Fig.~\ref{fig:spec}(d) and (f). We refrain from a quantitative comparison because ED is prone to finite size effects. By changing the simulated cluster geometry the gap between single-magnon states can vary considerably; the existence of the gap, however, is not subject to finite-size effects. (In Appendix \ref{sec:AppendixSpin1}, we use ED also to show that the Dirac magnons are gapped for $S=1$.)
We are thus content with concluding that spin-wave theory is a surprisingly good approximation even for $S=1/2$ and attribute this finding to the collinear texture and negligible frustration.

% ====================================
%  Discussion
% ====================================
\section{Discussion}
\label{sec:Discussion}
\subsection{Working principles for undamped interacting topological magnons}
\label{sec:workingprinciples}
From our results, we identify two working principles to minimize the damping caused by interactions without compromising their band-gap-inducing quality.

\paragraph*{(i) Tuning the Dirac point just below the lower threshold of the continuum.} To take full advantage of the logarithmic singularity in the real part of the self-energy (cf.~Sec.~\ref{sec:ResultsGap}) the Dirac cone must be energetically positioned close to but still outside of the continuum. This is easily realized by application of a magnetic field. Real materials may only be quasi-two-dimensional due to some (weak) interlayer coupling. Any degree of three-dimensionality regularizes the DOS of the two-magnon continuum and, hence, also the singular behavior of the self-energy. The logarithmic singularity and step of the real and imaginary part are replaced by a nonsingular square-root behavior \cite{Chernyshev2009}. Still, the gap-inducing real part is strongly enhanced at the threshold of the continuum.

\paragraph*{(ii) Artificially enhance interactions to expel the single-particle states from the continuum.} If principle (i) is not an option, e.g., because it would take inaccessibly large magnetic fields, one can try to make the continuum ``expel'' the single-particle states by artificially increasing interactions (cf.~Sec.~\ref{sec:ResultsRevival}). It is known that strong electric fields $\vec{E}$ cause DMI interaction because they break inversion symmetry \cite{Katsura2005, Liu2011Electric, Zhang2014Electric}. Hence, by applying an out-of-plane $\vec{E}$, which mimicks the electric field due to a structural potential gradient, Rashba-type DMI [as that considered in the chiral magnet in Fig.~\ref{fig:models}(b)] is enhanced. Moreover, recently, sereval ideas on light-induced effective magnetic interactions were discussed \cite{Struck2011, owerre2017floquet, Owerre2018photo, Elyasi2019, Malz2019, Bostrom2020}, which, when carefully designed, could cause considerable magnon-magnon interactions.

\subsection{Comparison between harmonic and interacting magnon Chern insulators}
\label{sec:harmonicVSanharmonic}
We reiterate that it is only due to the particle-number nonconserving interactions that the broken (effective) time-reversal symmetry of the chiral magnet in Fig.~\ref{fig:models}(b) becomes apparent (also cf.~Sec.~\ref{sec:Symmetry}). It is worth pointing out the difference to conventional magnon Chern insulators (e.g., Refs.~\onlinecite{Zhang2013, Owerre2016a}) that exhibit nontrivial topology already at the harmonic level. 
There, a reversal of the sign of the DMI flips the winding number and, hence, the edge state's chirality. This is because the harmonic theory sees $\vec{D} \cdot \vec{M}$, i.e., the projection of the DMI vectors $\vec{D}$ onto the magnetization $\vec{M}$. Hence, a reversal of $\vec{D}$ is equivalent to a reversal of $\vec{M}$, the latter being the actual crucial ingredient to flip the sign of topological invariants.
In contrast, the gap opening due to interactions observed in Sec.~\ref{sec:ResultsGap} is quadratic in DMI and, thus, independent of a sign flip of DMI. We argue that there is no necessity for the sign of DMI to be the relevant parameter. Instead, one verifies that it is still the magnetization reversal, which leads to a sign flip of the mass and, hence, also to a sign reversal of the winding number. To see this, we recall that upon magnetization reversal not all spin components acquire a minus sign but that the sign of one in-plane component stays invariant, e.g., for a rotation $\mathcal{R}(\vec{y},\pi)$, the sign of $S^y_{\vec{r}}$ stays invariant. Effectively, this is like flipping the sign of the $x$ components of the in-plane DMI vectors. Hence, the DMI-induced phases $\varphi_{\vec{\delta}_i} = \mathrm{arg}( d_{\vec{\delta}_i}^y - \mathrm{i} d_{\vec{\delta}_i}^x)$, which enter the interaction vertices via terms of the form $\mathrm{e}^{\mathrm{i} ( \varphi_{\vec{\delta}_i} - \vec{k} \cdot \vec{\delta}_i)}$ in Eqs.~\eqref{eq:V3a}-\eqref{eq:V3d}, are reversed, $\varphi_{\vec{\delta}_i} \to - \varphi_{\vec{\delta}_i}$. For the interaction vertices, such a flip amounts to the mapping $( V^{ \lambda, \mu \leftarrow \nu}_{\vec{k}, \vec{q} \leftarrow \vec{p}} )^\text{C} \to ( V^{ \lambda, \mu \leftarrow \nu}_{-\vec{k}, -\vec{q} \leftarrow -\vec{p}} )^{\text{C},\ast}$, i.e., to complex conjugation and reversal of crystal momentum, the two operations usually associated with time reversal. Hence, upon magnetization reversal, the interaction-induced mass term at the $K$ point is now found at the $K'$ point and vice versa. Obviously, the winding number in Eq.~\eqref{eq:winding} has to change sign.

One may also discuss what happens under a gradual rotation of $\vec{M}$ from up-pointing to down-pointing. Somewhere inbetween there has to be a topological phase transition associated with a gap closing. As $\vec{M}$ is fully rotated into the plane, the chiral magnet in Fig.~\ref{fig:models}(b) exhibits an effective time-reversal symmetry $\mathcal{T} C_{2,z}$. Taking a hexagon's center of mass as center for the $C_{2,z}$ rotation, which rotates both spins \emph{and} space (and, hence, maps the DMI vectors onto each other), actual time-reversal $\mathcal{T}$ maps the in-plane texture back onto itself. Hence, Dirac cones are stable, i.e., the band gap has closed.

\subsection{Relevance in real materials}
\label{sec:real-materials}
In real materials, magnonic topology is determined by a competition between harmonic and anharmonic mechanisms. For example, if we include a small $D_z$ second-nearest neighbor term [as considered in the achiral magnet, Fig.~\ref{fig:models}(a)] into the model of the chiral magnet [cf.~Fig.~\ref{fig:models}(b)], a Dirac mass and nontrivial topology are already found at the harmonic level \cite{Owerre2016a}, with the winding number determined by the sign of $D_z$. Additionally, sublattice-dependent on-site potentials, e.g., in the form of on-site anisotropies, may be considered. They break inversion symmetry and open a topologically trivial gap at the harmonic level. As is known from the electronic Haldane model \cite{Haldane1988}, a rich topological phase diagram with both topologically nontrivial and trivial phases is found at the single-particle level. For magnons, the effects of particle-number nonconserving interactions due to in-plane DMI ($D$) come on top. Since $D_z$ and $D$ can be independent---the former derives from spin-orbit coupling intrinsic to the material, while the latter is due to structural asymmetry, e.g., in the presence of a substrate---$D$ may be larger than $D_z$. Then, despite the $1/S$ smallness and the quadratic influence of $D$ on the dispersion, interactions may counteract and even undo or reverse harmonic topology, causing an opposite winding number or a transition from a topologically trivial into a nontrivial phase. 
With these effects being field as well as temperature-dependent, both magnon topology and transverse transport become highly nontrivial subjects. 
Since experimental studies of magnon topology and transverse transport have so far relied on the harmonic approximation for the interpretation of results, a careful reexamination may be necessary. 
Let us quickly go over selected materials for which there is experimental evidence compatible with nontrivial magnon topology and/or thermal Hall effects of magnonic origin. We restrict the discussion to (almost) collinear magnets.\footnote{For collinear magnets, the role of magnetic interactions in either harmonic ($S_i^\pm S_j^\pm$) or cubic Hamiltonians ($S_i^\pm S_j^z$) is clear, facilitating the discussion of their relative influence ($S_i^\pm = S_i^x \pm \mathrm{i} S_i^y$). Contrarily, for noncollinear and noncoplanar materials, a single type of magnetic interaction, say exchange interaction, is distributed over all sub-Hamiltonians and an analysis becomes much more complicated; a paradigmatic example is the Heisenberg antiferromanget on the triangular lattice \cite{Chernyshev2009}. It may very well be expected that anharmonic interactions in noncollinear magnets have an even larger influence on magnon topology and transverse transport than in collinear magnets.}

\paragraph*{Kagome ferromagnet Cu[1,3-benzenedicarboxylate(bdc)].}
The metal-organic kagome ferromagnet Cu(1,3-bdc), which exhibits Dirac gaps \cite{Chisnell2015} as well as transverse thermal transport \cite{Hirschberger2015} for an out-of-plane field, orders in-plane with (negligible) antiferromagnetic inter-plane coupling in the absence of an external field \cite{Chisnell2016}. Similar to the achiral model magnet in Fig.~\ref{fig:models}(a), the in-plane ordered state exhibits $D_z$-induced magnon damping \cite{Chernyshev2016}. However, also similar to the achiral magnet, $D_z$ does not break time-reversal symmetry, such that magnon-magnon interactions due to $D_z$ do not serve as an independent source of topology. However, note that since the kagome layers are no mirror planes in Cu(1,3-bdc), in-plane DMI components that span the entire plane are allowed \cite{Elhajal2002}. An in-plane ordered ferromagnetic state has finite projection onto some of the in-plane DMI vectors but is orthogonal to others. Hence, the time-reversal breaking influence of in-plane DMI enters both the harmonic as well as the cubic theory. An intricate interaction and/or counterplay is expected, influencing both magnon topology and the associated thermal Hall effect.\footnote{Note that a two-dimensional ferromagnet can exhibit a Chern insulating phase and transverse transport even if the out-of-plane magnetization is zero. The only requirement is that the in-plane magnetization direction breaks all crystalline symmetries that forbid out-of-plane pointing axial vectors. Hence, upon rotating the magnetization direction by $2\pi$ within the kagome planes of Cu(1,3-bdc) the thermal Hall conductivity should exhibit an oscillating pattern with period $2\pi/3$, similar to what was predicted for the electronic anomalous Hall effect in Ref.~\onlinecite{Liu2013QAHE} or for the thermal Hall effect in coplanar kagome antiferromagnets in Ref.~\onlinecite{Mook2019}. (The zeros of the oscillation are associated with magnetization directions coinciding with a mirror plane, installing a time-reversal mirror that forbids out-of-plane pointing axial vectors \cite{Neumann2020}.) Such a thermal Hall measurement may also clarify the magnitude of in-plane DMI vectors in Cu(1,3-bdc).}

Similarly, for the experimentally considered scenario of out-of-plane magnetized Cu(1,3-bdc), $D_z$ causes harmonic topology \cite{Katsura2010, Zhang2013, Mook14a, Mook14b} but the in-plane DMI vectors enter the cubic theory. Importantly, similar to the chiral model magnet in Fig.~\ref{fig:models}(b), the in-plane DMI vectors break the effective time-reversal symmetry even in the absence of $D_z$ and, hence, constitute an independent source of topology, which has been neglected so far \cite{Mook14a, Lee2015}. With the three-magnon interactions being easily ``field-frozen,'' a study of the magnetic-field dependence of topological magnonic gaps could clarify the relative importance of in-plane and out-of-plane DMI. Such results could shed new light on the origin of the experimentally observed sign changes in transverse transport \cite{Hirschberger2015}.

\paragraph*{Bilayer kagome antiferromagnet YMn$_6$Sn$_6$.}
The intermetallic room-temperature kagome antiferromagnet YMn$_6$Sn$_6$ exhibits both gapped magnon bands and Dirac magnons \cite{Zhang2020}. With the kagome lattice admitting of both in-plane as well as out-of-plane DMI vectors (the kagome planes are no mirror planes), any magnetization or N\'{e}el vector direction will select a subsection of DMI components entering the harmonic theory, with the remaining components appearing in the cubic Hamiltonian. Hence, with interactions and the free theory sharing symmetries, interactions should be considered for explaining neutron scattering data. Their effect may, however, be overshadowed by the rather large electron-magnon interaction causing Landau damping due to Stoner excitations \cite{Zhang2020}.

\paragraph*{Transition metal trihalides CrBr$_3$ and CrI$_3$.}
A temperature-driven many-body renormalization of Dirac magnons has been studied in the honeycomb-ferromagnet CrBr$_3$ \cite{Yelon1971}. The renormalization was successfully explained in terms of number-conserving four-particle interactions ($\hat{H}_4$), which derive from the exchange energy \cite{Pershoguba2018}, and, hence, do not gap out the Dirac magnons. The intrinsically small spin-orbit coupling (SOC) of CrBr$_3$ renders anisotropic spin-spin interactions irrelevant. However, one may think of growing CrBr$_3$ on a heavy-metal substrate to induce interfacial DMI for a realization of the chiral magnet model in Fig.~\ref{fig:models}(b).

The situation is drastically different in CrI$_3$, which exhibits mono-layer ferromagnetism due to SOC-induced anisotropies \cite{Huang2017}. To explain the experimentally observed gapped Dirac magnons \cite{Chen2018} two models---the $J$-$D_z$-model and the $J$-$K$-$\Gamma$-model\footnote{In the $J$-$D_z$-model, second-nearest neighbor DMI $D_z$ causes the topological gap opening, similar to Ref.~\cite{Owerre2016a}. In contrast, the $J$-$K$-$\Gamma$-model relies on the Kitaev interaction $K$ to open the gap, also topologically nontrivial \cite{McClarty2018,  Aguilera2020arxiv}.}---are debated \cite{Chen2020CrI3}, with the latter one having theoretical support \cite{Xu2018, Lee2020CrI3}.\footnote{However, note also that the magnon gap observed in experiments may not be global, if the Dirac cones are shifted in reciprocal space \cite{Ke2020}.} The anisotropic exchange interactions (sometimes called ``$\Gamma$-terms'') between nearest neighbors have a similar time-reversal symmetry breaking influence as the nearest-neighbor DMI in our chiral model magnet in Fig.~\ref{fig:models}(b). Hence, the associated three-magnon interactions may support or compete with the Kitaev terms that cause nontrivial magnon topology at the harmonic level. Furthermore, one may induce DMI by considering the intrinsically inversion asymmetric Janus monolayers \cite{Moaied2018}; recent theoretical studies have predicted very large DMI \cite{Xu2020, Yuan2020Janus, Liang2020}.

\paragraph*{Three-dimensional pyrochlore ferromagnet Lu$_2$V$_2$O$_7$ and multiferroic ferrimagnet Cu$_2$OSeO$_3$.}
Pyrochlore ferromagnets like Lu$_2$V$_2$O$_7$ exhibit a thermal Hall effect \cite{Onose2010, Ideue12} and were predicted to host Weyl magnons at high energies \cite{Mook2016, Su2016}. However, neutron scattering data revealed a rather blurred magnon spectrum \cite{Mena2014}, which may be brought about by spontaneous magnon decays due to the DMI vectors spanning the entire three-dimensional space. Hence, no matter in which direction the magnetization is pointing, DMI is distributed over harmonic as well as anharmonic Hamiltonians, rendering nonlinear spin-wave calculations necessary. 

Similar considerations apply to the Weyl magnon-hosting multiferroic ferrimagnet Cu$_2$OSeO$_3$ \cite{Zhang2020Weyl}. However, note that on top of DMI-derived three-magnon interactions there are also number-nonconserving four-magnon interactions (of type $b^\dagger b^\dagger b^\dagger b$) due to the \emph{ferri}magnetic spin alignment. Although they contribute to zero-temperature spontaneous decays only at order $1/S^2$, they are proportional to the exchange energy and do not suffer from the smallness of SOC. Hence, since $S=1/2$, the additional $1/S$ ``smallness'' does not compensate for the four-magnon vertices being larger by a factor of $J/D$ than the three-magnon vertices. Zero-temperature decays into the \textit{three}-magnon continuum are expected to be the dominating source of damping. 

Nonetheless, the Hermitian part of the interaction-induced self-energy due to DMI is expected to renormalize the position of the Weyl points. Moreover, the non-Hermitian part---so far only explored in two dimensions \cite{McClarty2019}---may introduce a new topological phase of magnons which is the magnonic analog of an ``exceptional topological insulator'' \cite{Denner2020}.

\paragraph*{Kamiokite Fe$_2$Mo$_3$O$_8$.}
The multiferroic kamiokite Fe$_2$Mo$_3$O$_8$ [and its derivatives (Zn$_x$Fe$_{1-x}$)$_2$Mo$_3$O$_8$] exhibits a giant thermal Hall effect \cite{Ideue2017}, which was argued to derive from phonon skew scattering (extrinsic contribution) and DMI-induced magnon-phonon interactions (intrinsic contribution) \cite{Park2020}; intrinsic contributions due to magnons were dismissed. The geometry between magnetization and DMI vectors is very much similar to the chiral model magnet in Fig.~\ref{fig:models}(b), and the arguments of Refs.~\onlinecite{Ideue2017, Park2020} for dismissing purely magnonic contributions rely on the accidental effective time-reversal symmetry of the harmonic theory. Our results challenge this conclusion; the in-plane DMI vectors of Fe$_2$Mo$_3$O$_8$ break the effective time-reversal symmetry via three-magnon interactions and will cause a \emph{purely} magnonic thermal Hall effect. Hence, a full theory of transverse thermal transport in kamiokite must consider magnons, phonons, the interactions among themselves and among each other.

% ====================================
%  Conclusion
% ====================================
\section{Conclusion}
\label{sec:Conclusion}
We demonstrated that magnon-magnon interactions may break a symmetry of the harmonic theory, admitting of a different topological phase as that found for non-interacting magnons. In particular, Dirac magnons in out-of-plane polarized honeycomb-lattice ferromagnets obtain a mass gap due to spontaneous quasiparticle decay, giving rise to a topogically nontrivial gapped phase with chiral edge states. Topological phase transitions brought about either by magnetic fields or temperature flip the chirality of the edge magnons, an experimental signature of which is a sign change in the thermal Hall conductivity.

These results show that magnon-magnon interactions do not only harbor detrimental lifetime broadening effects on the magnonic spectrum but constitute an origin of nontrivial topology on their own. Hence, they should not be neglected in studies of magnon topology. Potential material candidates to experimentally identify interaction-induced signatures either in the magnon spectrum or in transverse magnon transport are the kagome ferromagnet Cu(1,3-benzenedicarboxylate), the honeycomb-lattice van der Waals magnet CrI$_3$, and the multiferroic Fe$_2$Mo$_3$O$_8$.

Since we relied on a spin-to-boson transformation and extracted our results for the effective bosonic theory, the bosons may retrospectively be interpreted as different collective modes in solids. Hence, our general statements on the effect of interactions on single-particle topology also apply to, for example, phonons \cite{Zhang2010PhononHall}, triplons \cite{Romhanyi2015triplons, Joshi2019Z2triplon}, bosonic spinons (Schwinger bosons) \cite{Kim2016Schwinger}, magnon polarons \cite{Takahashi2016MagPhon, Zhang2020TopPhonon}, and magnon polaritons \cite{Okamoto2020}. Overall, our findings suggest that particle-number nonconserving many-body interactions are a crucial player in the field of topology of collective excitations in quantum condensed matter systems.

\begin{acknowledgments}
This work was supported by the Georg H.~Endress Foundation and the Swiss National Science Foundation and NCCR QSIT. This project received funding from the European Union’s Horizon 2020 research and innovation program (ERC Starting Grant, Grant No 757725). 
\end{acknowledgments}

\appendix

\section{Three-magnon interaction vertices for the achiral and chiral honeycomb ferromagnets}
\label{sec:Vertices}
In Eq.~\eqref{eq:H3HP}, we defined the Holstein-Primakoff three-boson vertices. For the chiral magnet, the only nonzero vertices are
\begin{subequations}
\begin{align}
    \left( V^{ 1,2 \leftarrow 1}_{\vec{k}, \vec{q} \leftarrow \vec{p}} \right)^\text{C} &= 
    - D \sqrt{\frac{S}{2}} \sum_{i=1}^3 \mathrm{e}^{\mathrm{i} ( \varphi_{\vec{\delta}_i} - \vec{q} \cdot \vec{\delta}_i)},
    \label{eq:V3a}
    \\
    \left( V^{ 2,1 \leftarrow 1}_{\vec{k}, \vec{q} \leftarrow \vec{p}} \right)^\text{C} &= 
    - D \sqrt{\frac{S}{2}}  \sum_{i=1}^3  \mathrm{e}^{\mathrm{i} ( \varphi_{\vec{\delta}_i} - \vec{k} \cdot \vec{\delta}_i)},
    \\
    \left( V^{ 2,1 \leftarrow 2}_{\vec{k}, \vec{q} \leftarrow \vec{p}}\right)^\text{C} &=  D \sqrt{\frac{S}{2}}  \sum_{i=1}^3 \mathrm{e}^{\mathrm{i} ( \varphi_{\vec{\delta}_i} + \vec{q} \cdot \vec{\delta}_i)},
    \\
    \left( V^{ 1,2 \leftarrow 2}_{\vec{k}, \vec{q} \leftarrow \vec{p}} \right)^\text{C} &=  D \sqrt{\frac{S}{2}}  \sum_{i=1}^3  \mathrm{e}^{\mathrm{i} ( \varphi_{\vec{\delta}_i} + \vec{k} \cdot \vec{\delta}_i)}, \label{eq:V3d}
\end{align}
\end{subequations}
where $\varphi_{\vec{\delta}_i} = \mathrm{arg}( d_{\vec{\delta}_i}^y - \mathrm{i} d_{\vec{\delta}_i}^x)$ is the ``phase'' of the DMI vector belonging to the nearest-neighbor bond $\vec{\delta}_i$ [see Eqs.~\eqref{eq:delta1}-\eqref{eq:delta3}]. Explicitly, the directions of DMI vectors read
\begin{subequations}
\begin{align}
	\vec{d}_{\vec{\delta}_1} &= (0,1), \\
	\vec{d}_{\vec{\delta}_2} &= (-\sqrt{3}/2,-1/2), \\
	\vec{d}_{\vec{\delta}_3} &= (\sqrt{3}/2,-1/2). 
\end{align}
\end{subequations}
For the achiral magnet, the nonzero vertices are
\begin{subequations}
\begin{align}
    \left( V^{ 1,1 \leftarrow 1}_{\vec{k}, \vec{q} \leftarrow \vec{p}} \right)^\text{A} &= 
    -\sqrt{2S} D_z \sum_{i=1}^3 \left[
    \sin\left( \vec{\tau}_i \cdot \vec{k}\right) + \sin\left( \vec{\tau}_i \cdot \vec{q}\right) \right]
    ,
    \\
    \left( V^{ 2,2 \leftarrow 2}_{\vec{k}, \vec{q} \leftarrow \vec{p}} \right)^\text{A} &= 
     \sqrt{2S} D_z \sum_{i=1}^3  \left[
    \sin\left( \vec{\tau}_i \cdot \vec{k}\right) + \sin\left( \vec{\tau}_i \cdot \vec{q}\right) \right]
    ,
\end{align}
\end{subequations}
where 
\begin{subequations}
\begin{align}
	\vec{\tau}_1 &= (0,-\sqrt{3}), \\
	\vec{\tau}_2 &= (3/2,\sqrt{3}/2), \\
	\vec{\tau}_3 &= (-3/2,\sqrt{3}/2),
\end{align}
\end{subequations}
connect second-nearest neighbors.

\section{Logarithmic singularities}
\label{sec:AppendixLogSingularities}
In Sec.~\ref{sec:ResultsGap}, we encountered a logarithmic singularity in the real part of the self-energy at the lower threshold of the two-magnon continuum. Here, we derive the analytic expression for this singularity.

We consider the kinematic situation that the Dirac energy $\varepsilon_\text{D}$ coincides with the lower threshold of the two-magnon continuum. This is the case for the critical magnetic field $B_\text{c} = JS$. The only decay channel for which the denominator of the self-energy in Eq.~\eqref{eq:self-energy-at-zeroT} can be resonant is that with both decay products in the lower branch: $j=j'=-$. The only possible energy and momentum-conserving decay channels for a Dirac magnon at $\vec{K}$ are those that involve two identical decay products at $\vec{K}/2$ (there are three such momenta, each halfway along the high-symmetry line connecting the $\varGamma$ point with the $K$ point; see also Sec.~\ref{sec:ResultsTransitionField}). Hence, we expand the field $B$ about $B_\text{c}$, and $\vec{q}$ about the two-magnon minimum at $\vec{K}/2$. In terms of
\begin{subequations}
\begin{align}
	\delta B &= B - B_\text{c},  \\
	\delta \vec{q} &= \vec{q} - \vec{K}/2,
\end{align}
\end{subequations}
the denominator of the self-energy becomes
\begin{align}
	\varepsilon_{\vec{K}} + \mathrm{i}0^+ - \varepsilon_{\vec{q},-} - \varepsilon_{\vec{K}-\vec{q},-}
	\approx
	\mathrm{i}0^+
	-
	\delta B 
	-
	\left( \frac{\delta q_x}{a} \right)^2
	-
	\left( \frac{\delta q_y}{b} \right)^2,
\end{align}
where $a$ and $b$ characterize the minimum of the two-magnon continuum along the $x$ and $y$ direction, respectively.

As far as the singular behavior of the self-energy is concerned, we may replace the interaction vertices by
\begin{align}
	\mathcal{V}^{\alpha \leftarrow --}_{ \vec{K}' \leftarrow \vec{K}'/2, \vec{K}'/2 }
    \mathcal{V}^{-- \leftarrow \beta}_{\vec{K}'/2, \vec{K}'/2 \leftarrow \vec{K}'}
    \to
    C^{\alpha\beta} D^2 ,
\end{align}
where we made the dependence on DMI explicit. Here, $C^{\alpha\beta}$ is a constant. For the achiral magnet $C^{\alpha\beta} = 0$ for $\alpha \ne \beta$, indicating the absence of an off-diagonal self-energy and, hence, of the gap. In contrast, for the chiral magnet $C^{\alpha \beta} \ne 0$ for all $\alpha$ and $\beta$.

The singular part of the self-energy thus reads
\begin{align}
	\varSigma^{\alpha \beta}_{\vec{K}}(\varepsilon_{\vec{K}})
	\sim
	C^{\alpha \beta} D^2   
    \int	
	\frac{
		\mathrm{d}^2 \delta q 
        }{
    \mathrm{i} 0^+ 
	-
	\delta B 
	-
	\left( \frac{\delta q_x}{a} \right)^2
	-
	\left( \frac{\delta q_y}{b} \right)^2
	},
	\label{eq:log-se}
\end{align}
and we extract
\begin{subequations}
\begin{align}
	\frac{\text{Re}\varSigma^{\alpha \beta}_{\vec{K}}(\varepsilon_{\vec{K}})}{J}
    &\sim
    C^{\alpha\beta} \left( \frac{D}{J} \right)^2 \log \left| \frac{\delta B}{J} \right|,
    \label{eq:appendix-real-sing}
    \\
    \frac{\text{Im}\varSigma^{\alpha \beta}_{\vec{K}}(\varepsilon_{\vec{K}})}{J}
    &\sim
    C^{\alpha\beta} \left( \frac{D}{J} \right)^2 \left[ \varTheta \left( \frac{\delta B}{J} \right) -1 \right],
    \label{eq:appendix-imag-sing}
\end{align}
\end{subequations}
from which Eqs.~\eqref{eq:real-singularity} and \eqref{eq:imag-singularity} follow. This singular behavior agrees with the general considerations of Ref.~\onlinecite{Chernyshev2009}.

\section{Off-shell solution of the Dyson equation}
\label{sec:OffShell}
To cut singularities in the renormalized magnon spectrum encountered within lowest-order perturbation theory, the authors of Ref.~\cite{Chernyshev2009} proposed a self-consistent off-shell solution of the Dyson equation. The main idea is to iteratively solve the Dyson equation, evaluating the self-energy at the so-obtained complex energies $\tilde{\varepsilon}_{\vec{k}}$, i.e.,
\begin{align}
	\tilde{\varepsilon}^{(\nu)}_{\vec{k}} = \varepsilon_{\vec{k}} - \mathrm{i} \eta + \varSigma_{\vec{k}}\left(\tilde{\varepsilon}_{\vec{k}}^{(\nu-1),\ast}\right), \label{eq:off-shell-Dyson}
\end{align}
where the complex conjugate $\tilde{\varepsilon}_{\vec{k}}^\ast$ accounts for causality \cite{Chernyshev2009}. $\nu$ denotes the iteration step. As an initial value one chooses $\tilde{\varepsilon}^{(0)}_{\vec{k}} = \varepsilon_{\vec{k}} - \mathrm{i} \eta $, with $\eta \ll \varepsilon_{\vec{k}}$ being a small numerical linewidth. Hence, after the first iteration one obtains the on-shell spectrum $\tilde{\varepsilon}^{(1)}_{\vec{k}}$. One then feeds $\tilde{\varepsilon}^{(1)}_{\vec{k}}$ back into the self-energy, calculates the new energies and repeats this process until $\tilde{\varepsilon}^{(\nu)}_{\vec{k}}$ converges. Since $\tilde{\varepsilon}^{(1)}_{\vec{k}}$ contains a finite imaginary part, i.e., a finite damping $\varGamma_{\vec{k}}$, the on-shell singularities in Eqs.~\eqref{eq:appendix-real-sing} and \eqref{eq:appendix-imag-sing} are cut. Effectively, the finite damping smears out the threshold of the two-magnon continuum, giving rise to a finite damping also for magnons below the continuum, that is, to those magnons that are nominally stable within the Born approximation. To see so, replace $\mathrm{i}0^+$ in Eq.~\eqref{eq:log-se} by $\mathrm{i}\varGamma_{\vec{k}}$.
Dropping $C^{\alpha\beta}$, and concentrating on the singularity at $\delta B = 0$, one obtains 
\begin{subequations}
\begin{align}
	\frac{\text{Re}\varSigma_{\vec{K}}(\varepsilon_{\vec{K}}+\mathrm{i}\varGamma_{\vec{K}})}{J}
    &\sim
    - \left( \frac{D}{J} \right)^2
    \log \left( 1 + \frac{\Lambda^2}{\varGamma_{\vec{K}}^2} \right),
    \label{eq:self-re}
    \\
    \frac{ \text{Im}\varSigma_{\vec{K}}(\varepsilon_{\vec{K}}+\mathrm{i}\varGamma_{\vec{K}}) }{J}
    &\sim
    - \left( \frac{D}{J} \right)^2 \arctan\left( \frac{\Lambda}{\varGamma_{\vec{K}}} \right),
    \label{eq:self-im}
\end{align}
\end{subequations}
where $\Lambda$ is an artificial ultraviolet integration cutoff. In principle, one would have to solve self-consistently both for the real and the imaginary part, accounting for an energy shift of the point where the single-particle energies cut into the continuum. For the sake of simplicity, we here only solve Eq.~\eqref{eq:self-im} self-consistently for the damping $\varGamma_{\vec{K}} = -\text{Im}\varSigma_{\vec{K}}(\varepsilon_{\vec{K}}+\mathrm{i}\varGamma_{\vec{K}})$ in the limit $\Lambda \gg \varGamma_{\vec{K}}$, resulting in $\varGamma_{\vec{K}} \sim D^2/J$. Thus, the damping is still perturbatively small but sufficient to cut the singularity of the real part in Eq.~\eqref{eq:self-re},
\begin{align}
	\frac{\text{Re}\varSigma_{\vec{K}}(\varepsilon_{\vec{K}}+\mathrm{i}\varGamma_{\vec{K}})}{J}
    &\sim
    - \left( \frac{D}{J} \right)^2
    \log \left( \frac{\Lambda J}{D^2} \right).
    \label{eq:real-off}
\end{align}
Hence, the real part exhibits a non-perturbative $D^2 \log |D|$ dependence. Both the damping and the real part of the self-energy vanish as $D \to 0$. The divergence as $\Lambda \to \infty$ in Eq.~\eqref{eq:real-off} is artificial because $\Lambda$ is physically bounded from above by the inverse of the lattice constant.

In writing Eq.~\eqref{eq:off-shell-Dyson}, we ignored to two-band nature of the honeycomb magnet. For the chiral magnet, there will be a spectral gap at $\vec{K}$ in the renormalized spectrum for $\nu \ge 1$. Hence, we decide to evaluate the new self-energy at the average energy, resulting in the equation
\begin{align}
	\tilde{\varepsilon}^{(\nu)}_{\vec{K},\pm} 
	= 
	\varepsilon_\text{D} - \mathrm{i} \eta 
	+ \varSigma^{--}_{\vec{K}}\left( \bar{\tilde{\varepsilon}}^{(\nu-1),\ast}_{\vec{K}} \right)
	\pm
	\varSigma^{+-}_{\vec{K}}\left( \bar{\tilde{\varepsilon}}^{(\nu-1),\ast}_{\vec{K}} \right),
	\label{eq:off-shell-Dyson-twoband}
\end{align}
with
\begin{align}
	\bar{\tilde{\varepsilon}}^{(\nu-1),\ast}_{\vec{K}} 
	=
	\frac{\tilde{\varepsilon}^{(\nu-1),\ast}_{\vec{K},+} + \tilde{\varepsilon}^{(\nu-1),\ast}_{\vec{K},-}}{2}.
\end{align}
Note that $\varSigma^{--}_{\vec{K}} = \varSigma^{++}_{\vec{K}}$ and $\varSigma^{+-}_{\vec{K}} = \varSigma^{-+}_{\vec{K}}$.

\begin{figure}
	\includegraphics[scale=1]{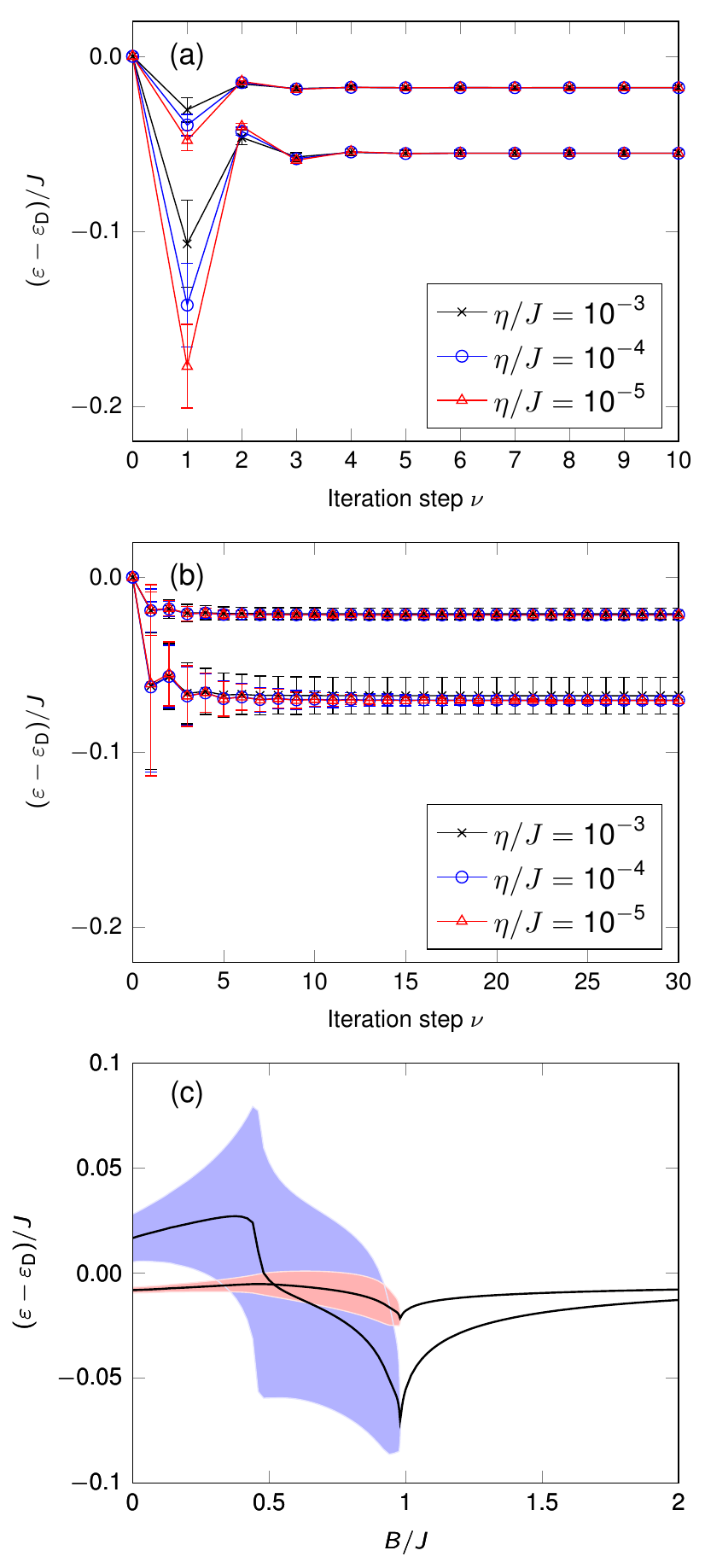}
	\caption{Magnon spectrum of the chiral magnet at the $\vec{K}$ point within the off-shell solution of the Dyson equation. 
	(a) Convergence behavior of the iterative off-shell solution at the singular point $B=B_\text{c}$ for selected values of the numerical linewidth $\eta$. No more than $10$ iteration steps are necessary to obtain converged results. Error bars indicate the damping.
	(b) Convergence behavior at $B/J = 0.981$. About $30$ iteration steps are necessary to obtain converged results for $\eta/J=10^{-5}$.
	(c) Magnon spectrum in dependence on magnetic field $B$ for $\eta/J = 10^{-5}$. Black lines indicate the real part of the spectrum and the colored regions indicate the linewidth (damping).
	Parameters read $S=1$ and $D/J=0.15$.} 
	\label{fig:offshell}
\end{figure}

First, we demonstrate the convergence behavior of Eq.~\eqref{eq:off-shell-Dyson-twoband} at the critical field $B = B_\text{c}$ for $S=1$ at $D/J=0.15$. Figure \ref{fig:offshell}(a) depicts the renormalized magnon energies as a function of the iteration step $\nu$. Error bars, given by $\pm \varGamma_{\vec{K},\pm}$, indicate the damping. At $\nu=0$, we show the harmonic magnon spectrum for which the Dirac cone is closed (zero mass). While the on-shell spectrum ($\nu=1$) exhibits the logarithmic divergence as $\eta \to 0$, the converged spectrum $\tilde{\varepsilon}^{(\nu \to \infty)}_{\vec{K},\pm}$ is independent of the numerical linewidth. We find that ten iteration steps are sufficient to ensure convergence at $B=B_\text{c}$. In contrast, more iteration steps (about $30$) are necessary to converge results for fields slightly below the critical field, as shown in Fig.~\ref{fig:offshell}(b) for $B/J = 0.981$. This is because the singularity is not only cut but also shifted towards lower energies.

Second, we explore the magnon spectrum at the $\vec{K}$ point in dependence on $B$, using $\eta/J = 10^{-5}$ and $30$ iteration steps. As shown in Fig.~\ref{fig:offshell}(b), the magnon spectrum of the chiral magnet is gapped, in general. For $B>B_\text{c}=J$, the two magnon branches are well-separated and have negligible damping. However, for $B<B_\text{c}$, decays into the two-magnon continuum cause strong lifetime broadening, as indicated by the colored areas. With the damping being larger than the gap, the notion of the gap ceases to exist. Interestingly, the gap becomes larger than the damping again for small fields. Overall, the off-shell solution for the magnon spectrum agrees very well with the spectral function in Fig.~\ref{fig:spec}(f). In particular, there is no divergence at $B = B_\text{c}$ but merely a cusp.

\section{Details of slab calculations for the chiral magnet}
\label{sec:AppendixDetailsSlab}
\begin{figure}
	\centering
	\includegraphics[scale=1]{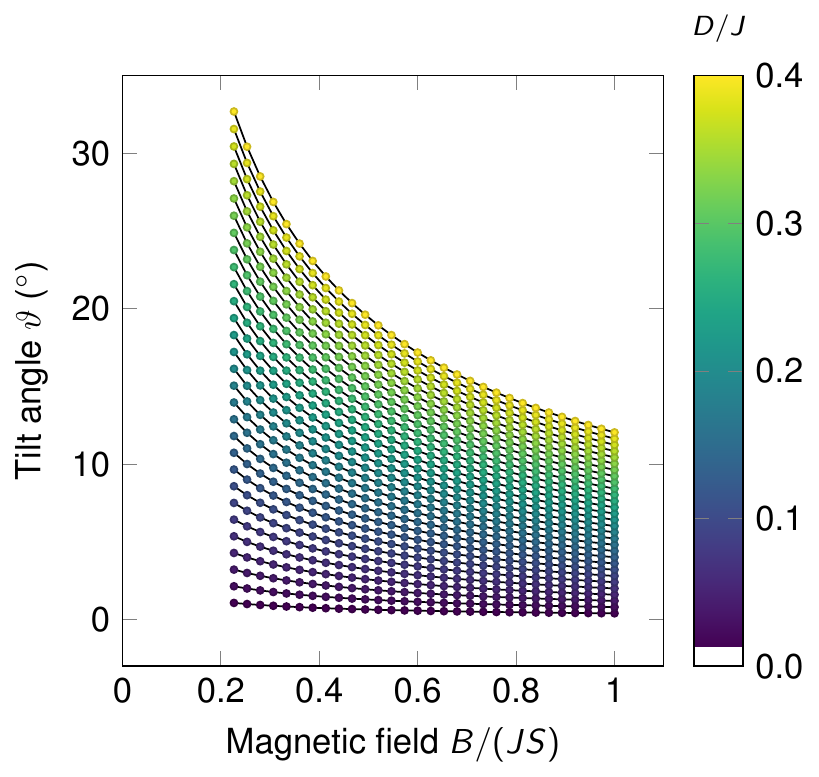}
	\caption{DMI-induced edge twist quantified by the edge tilt angle $\vartheta$ in dependence on magnetic field $B$ and $D/J$.}
	\label{fig:tilt-angle}
\end{figure}

Finite samples (slabs) feature boundary spins that miss neighbors. Hence, the chiral DMI is no longer compensated and there is an edge-located relaxation of the magnetic texture away from the collinear state. This phenomenon is known as ``surface twist'' and its amplitude decays exponentially towards the bulk \cite{Meynell2017}. To quantify this effect, we study the tilt angle $\vartheta$ of the boundary spins in dependence on DMI and magnetic field, as obtained from a classical energy minimization procedure based on an overdamped Landau-Lifshitz-Gilbert equation. The results in Fig.~\ref{fig:tilt-angle} show that for $D/J=0.2$ the tilt decreases from $\vartheta \approx 16^\circ$ to $6^\circ$ as the field increases from $B/(JS) \approx 0.2$ to $B/(JS) \approx 1$. Upon rotating slightly into the plane, the edge spins exhibit a finite projection onto the in-plane DM vectors, hence contributing to the harmonic theory which leads to nonreciprocal features of boundary-located magnons ($\varepsilon_{\vec{k}} \ne \varepsilon_{-\vec{k}}$), as expected for a locally broken inversion symmetry. It does not, however, lead to a band gap opening on the harmonic level, because the bulk magnetization is still pointing out of the plane. Hence, we neglect the boundary twist altogether and consider the texture to be strictly collinear. (Doing so, we also neglect the renormalizing effect of many-body interactions on the ground state directions of edge spins.)

The reduced coordination of edge spins has another effect: edge excitations are shifted downwards in energy relative to bulk excitations. Colloquially speaking, since edge spins are ``more floppy'' than bulk spins, it does not cost as much energy to excite them. For our study, this becomes relevant when analyzing edge states. For example, a zigzag-terminated edge of the electronic honeycomb lattice is known to feature a flat state, which connects the projections of the two Dirac cones \cite{Nakada1996, Yao2009GrapheneEdge, Brey2006, Lado2015Graphene}. For magnons, however, this state is no longer flat but looks like a loose drumhead \cite{Pantaleon2018}. Such phenomena have been observed, e.g., for the drumhead surface states associated with magnonic nodal lines \cite{Mook2016b, Pershoguba2018}. Although this is in principle no major road block, it increases the energy window covered by the edge states. With the slab calculations being computationally demanding, any reduction of the relevant energy window is facilitating the analysis. Hence, we apply local magnetic fields only to the edge spins. (Physically, this situation mimicks proximitizing the magnet with another magnet.) The magnitude of the local field is chosen to exactly compensate for the effective field of the missing neighbors. Hence, without interactions, the magnonic edge states of a zigzag-terminated honeycomb ferromagnet become flat akin to their electronic analogs. We reiterate that the local field by itself has no influence on topology and just shifts the edges states in energy. Moreover, the edge field also decreases the effect of the surface twist even more, providing an additional reason to neglect any twists. 

With the necessary approximations established, we continue with considering a slab with periodic boundary conditions along the $y$ direction but open boundary conditions in the $x$ direction. Hence, the edges are zigzag-terminated. We choose a width of $12$ honeycomb-lattice unit cells, resulting in a slab supercell of $24$ spins. The harmonic Hamiltonian 
\begin{align}
	\hat{H}_2^\text{slab} = \sum_{\vec{k}} \hat{\vec{A}}^\dagger_{\vec{k}} \cdot \mathcal{H}^\text{slab}_{\vec{k}} \cdot \hat{\vec{A}}_{\vec{k}},
\end{align} 
with 
$
    \hat{\vec{A}}_{\vec{k}}^\dagger = (\hat{a}_{\vec{k},1}, \ldots, \hat{a}_{\vec{k},24})
$ 
and a $24$-by-$24$ Hamilton kernel $\mathcal{H}^\text{slab}_{\vec{k}}$, is readily constructed numerically by a straightforward application of linear spin-wave theory. It is diagonalized by the matrix $\mathcal{U}_{\vec{k}}^\text{slab}$. 

We also use numerics to construct the three-magnon interaction vertices between the $24$ types of magnon normal modes. We rely on Refs.~\onlinecite{McClarty2018} and \onlinecite{Mook2020SkyrmionDamping} for the general expressions of the vertices. The numerical calculation of the self-energy $\varSigma^\text{slab}_{\vec{k}}(\varepsilon)$ and Green's function $\mathcal{G}^\text{slab}_{\vec{k}}(\varepsilon)$ follow the standard procedure and are done in the eigenbasis. To facilitate numerical efforts, we only evaluate the tridiagonal entries of $\varSigma^\text{slab}_{\vec{k}}(\varepsilon)$ (main diagonal and the first diagonals above and below it) because those are sufficient to capture band gap openings between degenerate bands to order $1/S$.

To obtain a spatially resolved spectral function 
\begin{align}
	A^\text{slab}_{i,k_y}(\varepsilon) = -\frac{1}{\pi} \text{Im} \left[ \mathcal{U}_{\vec{k}}^\text{slab} \mathcal{G}^\text{slab}_{\vec{k}}(\varepsilon) \mathcal{U}_{\vec{k}}^{\text{slab},\dagger} \right]_{i,i},
\end{align}
where $i=1, \ldots, 24$ labels a site of the slab unit cell ($i=1$ and $i=24$ label the left-most and right-most spin, respectively), we transform the Green's function back to the Holstein-Primakoff basis. Then, we define
\begin{subequations}
\begin{align}
	A^\text{left}_{k_y}(\varepsilon) = \sum_{i=1}^{12} A^\text{slab}_{i,k_y}(\varepsilon), \\
	A^\text{right}_{k_y}(\varepsilon) = \sum_{i=13}^{24} A^\text{slab}_{i,k_y}(\varepsilon)
\end{align}
\end{subequations}
and plot $A^\text{left}_{k_y}(\varepsilon) - A^\text{right}_{k_y}(\varepsilon)$ in Fig.~\ref{fig:slab}(b), choosing blue/red color for positive/negative values.

\section{Exact diagonalization for $S=1$}
\label{sec:AppendixSpin1}
\begin{figure}
	\includegraphics[scale=1]{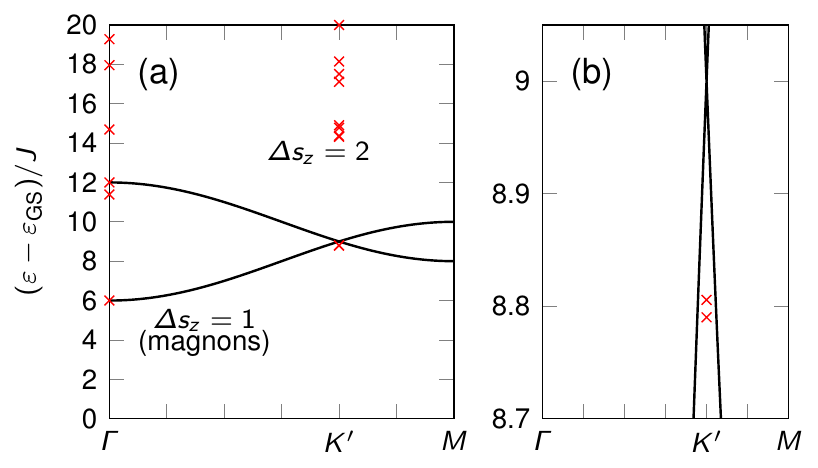}
	\caption{Spectrum of the chiral magnet obtained from exact diagonalization for $S=1$. Energies are given with respect to the ground state energy $\varepsilon_\text{GS}$. (a) Momentum dependence of the spectrum for $D/J=0.15$ at $B/J = 6$, causing an energetic separation between $\Delta s_z = 1$ excitations (magnons) and higher-spin excitations. Black lines indicate the spin-wave spectrum given in Eq.~\eqref{eq:single-particle-energy}. (b) Zoom into the relevant energy window of Dirac magnons, which are clearly gapped. A cluster of $18$ spins was considered.} 
	\label{fig:EDspin1}
\end{figure}

For $S=1$, ED of a $9$-unit cell honeycomb flake with periodic boundary conditions is considered ($18$ spins in total). Figure \ref{fig:EDspin1}(a) shows a comparison between the results of ED (red marks) and the spin-wave dispersion [Eq.~\eqref{eq:single-particle-energy} for $S=1$] in the limit of large fields that cause a separation between $\Delta s_z=1$ and higher-spin excitations. However, the field $B/J = 6$ is still too small to fully separate one-magnon from two-magnon excitations, causing an overlap of the two sectors at the $\varGamma$ point. Since DMI is nonzero ($D/J=0.15$), the Dirac magnons at the $K$ point get downwards renormalized and gapped [see Fig.~\ref{fig:EDspin1}(b)]. Hence, the spectra obtained within ED for both $S=1/2$ as well as $S=1$ exhibit the same qualitative features as those calculated from nonlinear spin-wave theory.

\newpage

\bibliographystyle{naturemag}
\bibliography{newrefs}

\end{document}